\documentstyle[epsfig,axodraw]{iop}          

\def\bear{\be\begin{array}}      
\def\eear{\end{array}\ee}
\def\bea{\begin{eqnarray}}
\def\eea{\end{eqnarray}}
\def\ie{{\it i.e.}}
\def\etal{{\it et al.}}

\def\bold#1{\setbox0=\hbox{$#1$}%
     \kern-.025em\copy0\kern-\wd0
     \kern.05em\copy0\kern-\wd0
     \kern-.025em\raise.0433em\box0 }
\def\beq{\begin{equation}}
\def\eeq{\end{equation}}
\def\bea{\begin{eqnarray}}
\def\eea{\end{eqnarray}}
\def\bq{\begin{quote}}
\def\eq{\end{quote}}

\def\IJMP{{\it Int.J.Mod.Phys.} }

\def\MPL{{\it Mod.Phys.Lett.} }

\def\NC{{\it Nuovo Cimento} }
\def\NP{{\it Nucl.Phys.} }
\def\NPPS{{\it Nucl.Phys.Proc.Supp.} }
\def\PL{{\it Phys.Lett.} }

\def\PR{{\it Phys.Rev.} }
\def\PRL{{\it Phys.Rev.Lett.} }
\def\PRTS{{\it Physics Reports} }

\def\PTP{{\it Progr.Theor.Phys.} }

\def\RNC{{\it Rivista del Nuovo Cimento} }

\def\ZP{{\it Z.Phys.} }

\def\gappeq{\mathrel{\rlap {\raise.5ex\hbox{$>$}}
{\lower.5ex\hbox{$\sim$}}}}

\def\lappeq{\mathrel{\rlap{\raise.5ex\hbox{$<$}}
{\lower.5ex\hbox{$\sim$}}}}

\def\DESepsf(#1 width #2){\epsfxsize=#2 \epsfbox{#1}}
 
\newcommand{\rpv}{{\not \!\! R_p}}
\newcommand{\gsim}{\stackrel{>}{\sim}}
\newcommand{\lessim}{\stackrel{<}{\sim}}

\def\xi{\ensuremath{\chi^{+}}}

\newcommand{\LEPONE}{\mbox{\sc LEP 1~}}
\newcommand{\LEPTWO}{\mbox{\sc LEP 2~}}
\newcommand{\LEPTWOHUN}{\mbox{\sc LEP 200~}}

\def\pb{\;{\rm pb}}
\def\lapproxeq {\mbox{{\lower .7ex\hbox{$\;\stackrel{\textstyle
                  <}{\sim}\;$}}}}
\def\gapproxeq  {\mbox{{\lower .7ex\hbox{$\;\stackrel{\textstyle
                  >}{\sim}\;$}}}}

\newcommand{\MW}{\mbox{$M_{\mathrm{W}}$}}

\newcommand{\ee}{\mbox{$\mathrm{e}^+\mathrm{e}^-$}}
\newcommand\new{\newcommand}         
\catcode`\"=\active                  
\new"[1]{{\accent127#1}}             
\new\3{\ss }                         
\new{\mm}[1]{{\mbox{\hspace{#1mm}}}} 

\new\Tab[1]{Table~\ref{tab:#1}}

\new\dmw{\mbox{$\Delta \MW$}}

\newcommand{\ALEPH}{\mbox{\sc ALEPH}}
\newcommand{\DELPHI}{\mbox{\sc DELPHI}}
\newcommand{\OPAL}{\mbox{\sc OPAL}}

\newcommand{\LEP}{\mbox{\sc LEP~}}
\newcommand{\ZEUS}{\mbox{\sc ZEUS~}}
\newcommand{\HONE}{\mbox{\sc H1~}}
\newcommand{\HERA}{\mbox{\sc HERA~}}
\newcommand{\CDF}{\mbox{\sc CDF~}}
\newcommand{\LHC}{\mbox{\sc LHC~}}
\newcommand{\ATLAS}{\mbox{\sc ATLAS~}}
\newcommand{\CMS}{\mbox{\sc CMS~}}

\def \gsim{\stackrel{\scriptstyle {>}}{\scriptstyle {\sim}}}

\def\sba2{\sin ^2 (\beta - \alpha)}
\def\cba2{\cos ^2 (\beta - \alpha)}
\def\tanB{\tan \beta}
\def\r{\rightarrow}
\def\Ecm{E_{CM}}
\def\ss{\sqrt{s}}
\def\h{\mathrm h}
\def\A{\mathrm A}

\def\Z{\mathrm Z}

\def\HH{\mathrm H}
\def\Hpm{\mathrm H^\pm}

\def\ee{\mathrm e^+\mathrm e^-}
\def\mm{\mu^{+}\mu^{-}}

\def\pb{ \mathrm{pb} ^{-1}}

\def\G{\mathrm{GeV}}

\def\bb{\mathrm b \bar{\mathrm b}}

\def\mh{m_{\h}}

\def\mHH{m_{\HH}}
\def\mZ{m_{\Z}}

\def\mt{m_{\mathrm t}}

\def\G{ \rm{GeV} }

\newcommand{\smu}{\tilde{\mu}}

\newcommand{\msmu}{\mbox{$m_{\smu}$}}
\newcommand{\mchz}{\mbox{$m_{\tilde{\chi}^0_1}$}}
\newcommand{\ipb}{\mbox{pb$^{-1}$}}
\newcommand{\IL}{\mbox{$\cal L$}}
\newcommand{\nt}{\tilde{\chi}^0}
\newcommand{\smp}{Standard Model processes}

\begin{document} 
\begin{flushright}
RAL TR-97-037 \\ hep-ph/9708250 \\
\end{flushright}
\title{Report of the 1997 LEP2 Phenomenology Working Group on `Searches'
(Oxford)\footnote{To appear in Journal of Physics G}}[Searches]
 
\author{B~C~Allanach$^1$, G~A~Blair$^2$$^\ast$, 
M~A~Diaz$^{11}$, 
H~Dreiner$^1$$^\ast$, 
J~Ellis$^3$, P~Igo-Kemenes$^4$, S~F~King$^5$ , P~Morawitz$^6$, W~Murray$^1$, 
A~Normand$^7$, D~A~Ross$^5$, P~Teixeira-Dias$^8$, 
M~D~Williams$^6$, G~W~Wilson$^9$, T~Wyatt$^{10}$}

\address{$^1$Rutherford Appleton Laboratory, UK.\\
$^2$Royal Holloway, Univerity of London, Egham, UK.\\
$^3$Theoretical Physics Division, CERN, Geneva, Switzerland\\
$^4$Physikalisches Institut, University of Heidelberg, FRG.\\
$^5$Physics Department, University of Southampton,\\
Southampton, SO17 1BJ, U.K.\\
$^6$Dept. of Physics, Imperial College, London, UK.\\
$^7$Dept. of Physics, University of Liverpool, UK.\\
$^8$Dept. of Physics and Astronomy, University of Glasgow, UK.\\
$^9$DESY/U. Hamburg, FRG\\
$^{10}$Dept. of Physics, University of Manchester, U.K.\\
$^{11}$Departamento de Fisica Teorica, Universidad de Valencia, Burjassot,
Valencia 46100, Spain}

\address{$^*$ convenors}

\begin{abstract} 
The Searches Working Group discussed a variety of topics 
relating to present and future measurements of searches at \LEPTWO. 
The individual contributions are included separately.
\end{abstract} 
 
\section{Introduction}

The `Searches' working group addressed the prospects for searches
for supersymmetry, higgs boson production and leptoquark production
at \LEPTWO.  The present status of Higgs searches, SUSY searches
and Supergravity were well covered in the plenary talks by
P~Igo-Kemenes, S~Katsanevas and G~G~Ross in these proceedings.  

The working group met together for seminar presentations which set the
agenda for the main themes of study.  One of the working
subgroups subsequently formed to concentrate on the issues
of higgs production, where realistic prospects for discovery or
for setting mass limits are presented in section 2.  Other subgroups
addressed prospects for supersymmetry through theoretical studies of 
radiative corrections to chargino production, section 3, constraints on
model parameters arising from colour and charge breaking, section 4, and 
theoretical issues relating to R-parity non-conservation, section 5.
Also addressed were the experimental prospects for supersymmetry 
in the contexts of slepton production, section 6
and R-parity violating production of single sneutrinos in section 7.
The renewed interest in leptoquark searches was motivated by recent reports 
from the \HERA experiments
and a summary is included below, in section 8.
The working group contribution is summarised by J~Ellis in the final section.


\section{Prospective sensitivity of Higgs boson searches at \LEPTWO  in
  1997 and beyond}

\author{P~Igo-Kemenes, W~Murray, A~Normand and P~Teixeira-Dias}

\begin{abstract}

  A reassessment of the luminosity and centre-of-mass energy
  conditions needed to exclude/discover a Higgs boson signal at \LEPTWO 
  is presented, both for the neutral Higgs boson predicted in the 
  context of the Standard Model and for the lightest neutral Higgs boson
  of the Minimal Supersymmetric Standard Model.  This reconsideration is 
  based on the results prepared before the start of \LEPTWO  and on recent
  studies incorporating more up-to-date knowledge of the performance of 
  the four \LEP experiments at \LEPTWO  energies.

\end{abstract}

\subsection{Introduction}
The search for Higgs bosons is one of the major topics being pursued
at \LEPTWO.  Since the end of the first phase of \LEP, which operated from 
1989 to 1995 at energies around the $\Z$ resonance, the collider has
run at centre-of-mass energies of 130--136, 161 and 172,$\G$.

So far, no evidence of Higgs particles has been found and the four
\LEP experiments have set lower limits, at the 95\% confidence level 
(C.L.), on the mass of a Standard Model (SM) Higgs boson in the range
65--71\,GeV/$c^2$ \cite{SM172}.  A preliminary estimate of the combined
result of the four \LEP experiments yields $\mHH >$ 77\,GeV/$c^2$ 
(95\% C.L.) \cite{privcomm}.

In 1997 \LEP\ will operate at $\ss=183\,\G$ and a further increase
of around 10\,$\G$ in the centre-of-mass energy is expected in 1998.
Here we consider the prospects for experimentally excluding or
discovering a Higgs boson at \LEPTWO  in 1997 and beyond.  Results
are presented in terms of the two relevant collider parameters, the
average integrated luminosity delivered to the experiments, $\cal L$,
and the centre-of-mass energy, $\ss$.  We first examine the case of
the SM Higgs boson and then consider the Higgs sector of the Minimal
Supersymmetric Standard Model (MSSM).

\subsection{Standard Model Higgs boson}

The results presented here are based on  

\begin{itemize}
\item[(a)] the studies, prepared before the start of \LEPTWO, of
  the integrated luminosity needed to exclude or to discover a Higgs
  boson signal with a given mass, $\mHH$, at centre-of-mass energies of
  175, 192 and 205\,$\G$ \cite{YellowBook};
\item[(b)] a similar, but more recent, study prepared for the 1997 \LEP\
  Performance Workshop \cite{Chamonix97}, incorporating new information 
  from the four \LEP\ experiments based on the analysis of the first data
  collected at \LEPTWO  energies, extrapolated to 185 and 189\,$\G$.
\end{itemize}

The comparison of (a) and (b) shows that the Higgs search sensitivity
of the \LEP\ experiments has improved from the predictions in
reference \cite{YellowBook}.  Quantitatively we note that, for given
$\cal L$ and $\ss$ conditions, the Higgs boson mass sensitivity in (b)
is better by, on average, approximately $1.2$\,GeV/$c^2$ than in (a). The
values in the latter have therefore to be corrected as 

\[ \mHH^{\mathrm (b)} = \mHH^{\mathrm (a)} + 1.2\,{\mathrm GeV}/c^2, \]

\noindent except that $\mHH^{\mathrm (b)}$ is constrained not to exceed 
the previous mass limit for a given $\ss$ at high luminosity.
This improvement is largely due to the effort invested by the
experimental collaborations in increasing their sensitivity to a
possible Higgs boson signal by using, for example, more sophisticated
b-tagging methods and optimised selections.

\begin{figure}[hbtp]
\begin{center}
\begin{tabular}{cc}
\mbox{\epsfig{file=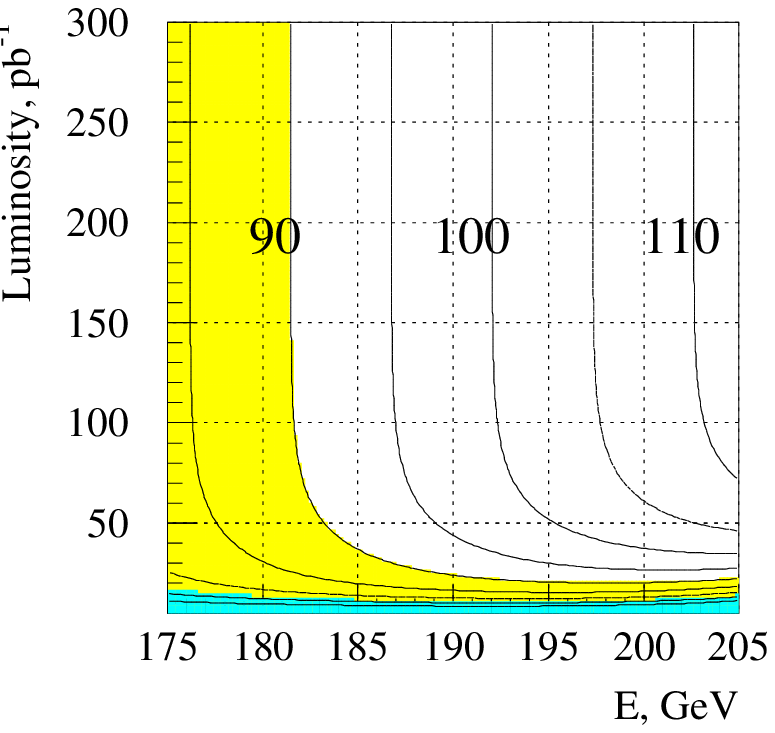,height=6cm,bbllx=0,bblly=0,bburx=220,
bbury=220}} &                         
\mbox{\epsfig{file=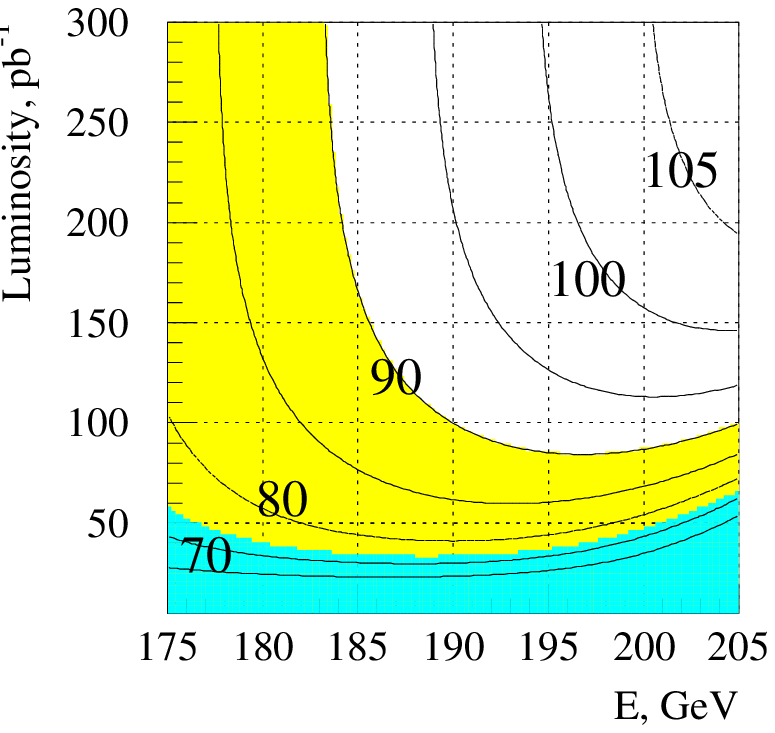,height=6cm,bbllx=0,bblly=0,bburx=220,
bbury=220}} 
\end{tabular}
\end{center}
\caption{
  (a) The integrated luminosity per experiment and centre-of-mass energy
  conditions required to exclude, at the 95\% C.L., the existence of a SM
  Higgs boson of given mass, as indicated by the contours, with the four  
  \LEP\ experiments combined; (b) same as in (a) for a $5\sigma$ discovery. 
  The dark and light shaded areas indicate the anticipated 95\% C.L. 
  {\bf excluded} regions after the 1996 and 1997 data-taking respectively,
  assuming 100\,$pb$ per experiment in 1997.}
\label{fig:SM}
\end{figure}

Figure \ref{fig:SM}(a) shows iso-sensitivity contours for exclusion at 
the 95\% C.L. of a Higgs boson of given mass, projected on to the ${\cal
L}$--$\ss$ plane.  From this it can be concluded that the 95\% C.L. 
exclusion limit is driven by the centre-of-mass energy provided that more
than a certain luminosity, ${\cal L_{\mathrm threshold}} \simeq 50\,\pb$,
is delivered to each experiment.  Above this threshold the excluded Higgs 
boson mass varies essentially linearly with $\ss$:

\[ \mHH^{\mathrm{excl}} = (\ss - 91)\,{\mathrm GeV}/c^2.\]

%
%

The $5\sigma$ discovery potential, as defined in \cite{YellowBook},
for the SM Higgs boson is shown in figure \ref{fig:SM}(b).
For instance, for the discovery of a Higgs particle with mass $\mHH \simeq
\mZ$ there is a slight preference for a centre-of-mass energy
$\ss \simeq 195\,\G$.  While such a signal could be discovered at this
energy with approximately 85\,$\pb$ per experiment, a larger
luminosity of around 100\,$\pb$ per experiment would be required at 
$\ss = 205\,\G$.  This can be explained by the behaviour with $\ss$ of the
signal and effective background cross-sections.  For $\mHH \simeq \mZ$,
the effective background (i.e. the background remaining after
the experimental event selections) is dominated by the irreducible
$\Z\Z$ background.  Table \ref{tab:hzzz} shows the contributions to the
cross-section for the $\ee\r\bb\mm$ process, for example, obtained by
calculating the $\HH\Z$ and $\Z\Z$ diagrams separately, and the total
calculated cross-section.  It can be seen from this table that the ratio 
of signal to background cross-sections falls continuously above $\ss 
\simeq 185\,\G$, thereby increasing the integrated luminosity that would
be required to discover a Higgs boson of mass $\mHH \simeq \mZ$.
\begin{table}[hbtp]
  \caption{Contributions to the cross-section for the process $\ee\r\bb\mm$
    obtained by calculating the $\HH\Z$ and $\Z\Z$ diagrams separately when
    $\mHH$ = 90\,GeV/$c^2$, and the total calculated cross-section.  Values 
    are obtained from the WPHACT1.0 generator \cite{WPHACT} with cuts ($50<
    m_{\bb}<300$)\,GeV/$c^2$ and ($\mZ-25<m_{\mm}<\mZ+25$)\,GeV/$c^2$.}
  \label{tab:hzzz}
  \lineup
  \begin{indented}
    \item[]\begin{tabular}{@{}lllllll}                                      \br
      $\ss$ (GeV)           &  175 &  185  &  192  &  195  &  200 &  205 \\ \mr
      $\sigma_{\HH\Z}$ (fb) & 0.39 &\06.66 & 10.6  & 11.5  & 12.3 & 12.6 \\ 
      $\sigma_{\Z\Z}$  (fb) & 0.82 &\04.81 &\08.65 &\09.65 & 10.8 & 11.5 \\ 
      $\sigma_{total}$ (fb) & 1..20 & 11.5  & 19.3  & 21.1  & 23.1 & 24.1 \\ \br
    \end{tabular}
  \end{indented}
\end{table}

During 1997, \LEPTWO  will operate at $\ss = 183\,\G$ and it is expected
that a luminosity of approximately 100\,$\pb$ will be delivered to each 
of the experiments.  Under these conditions, the combined sensitivity of 
the four \LEP\ experiments to the SM Higgs boson after the 1997 run would
reach $\mHH \simeq \mZ$ for 95\% C.L. exclusion and $\mHH \simeq$
85\,GeV/$c^2$ for the case of a $5\sigma$ discovery \cite{privcomm}.

\subsection{MSSM Higgs sector}

In the Higgs sector of the MSSM, two complex doublets of scalar fields are
needed with vacuum expectation values (VEVs) $v_1$ and $v_2$, coupling to
down-type and up-type fermions respectively.  One of the parameters of the
model is the VEV ratio $\tanB=v_2/v_1$.  The Higgs spectrum consists of 
five mass eigenstates, $\h$, $\A$, $\HH$ and $\Hpm$.  The lightest CP-even 
neutral scalar, $\h$, and the CP-odd neutral scalar, $\A$, may be detected 
at \LEPTWO. 

\begin{figure}[hbtp]
\centerline{\epsfig{file=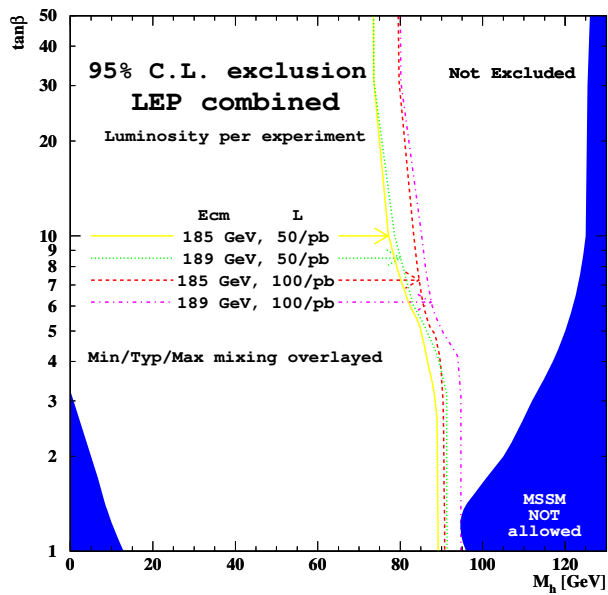,width=8.0cm}}
\caption{95\% C.L. exclusion sensitivity in the MSSM with four \LEP\
  experiments combined. The theoretically allowed region is defined as 
  the set of all $(\mh,\tanB)$ points allowed by at least one of the 
  three (minimal, typical, maximal) stop mixing scenarios.  (Figure 
  supplied by the Joint \LEP\ Higgs Working Group.)}
\label{fig:MSSM}
\end{figure}

Excluded areas of the $(\mh,\tanB)$ plane are shown in figure 
\ref{fig:MSSM} \cite{Chamonix97}.  Unphysical regions excluded by all of 
the three benchmark stop mixing configurations (minimal, typical and 
maximal mixing \cite{YellowBook}) are indicated in black. The projected 
95\% C.L. excluded areas are shown for integrated luminosities of 
50\,$\pb$ and 100\,$\pb$ per experiment.

The experimental sensitivity in the {\bf high} $\tanB$ ($\gsim 5$)
region is dominated by searches for the pair-production process
$\ee\r\h\A$.  From figure \ref{fig:MSSM} it can be seen that the
exclusion there is essentially driven by the accumulated luminosity 
rather than the centre-of-mass energy.  Although \LEPTWO  is expected to 
improve the excluded region for high $\tanB$, it will not do enough to 
bridge the gap between the currently excluded and theoretically forbidden 
regions. This will be an important part of the \LHC research programme.

The {\bf low} $\tanB$ region is covered by searches for the
Higgs-strahlung process.  SM searches for $\ee\r\HH\Z$ can
be interpreted as MSSM searches for $\ee\r\h\Z$, albeit with lower
sensitivity because of the additional factor $\sba2$ in the
cross-section for the latter, where $\alpha$ is a mixing angle in the 
CP-even Higgs sector.  Typically, the exclusion limits for the $\h$ boson 
will be 1--2\,$\G$ lower than for the SM Higgs boson.  

The expected \LEP\ running conditions of 1997 will allow the experiments 
to probe the low $\tanB$ range almost to the upper bound on $\mh$. This 
is interesting in particular in the context of the MSSM b--$\tau$ Yukawa
coupling unification scenario \cite{YellowBook,b-tau}.  In this scenario 
only the ranges $1 \lessim \tanB \lessim 3$ and $\tanB \gsim 50$ are allowed, 
other $\tanB$ values being incompatible with current measurements of the 
top quark mass \cite{mtop}.

The first of these $\tanB$ ranges can be probed at \LEPTWO  if sensitivity 
to $\mh$ between 95 and 112\,GeV/$c^2$ is achieved (figure \ref{fig:MSSM}).
An $\h$ boson with $\mh=95$\,GeV/$c^2$ can be excluded at the 95\% C.L. by 
the four \LEP\ experiments combined if, for example, $\ss =$ 188--189\,$\G$  
and an integrated luminosity of 100\,$\pb$ per experiment is obtained.  
>From the SM Higgs boson results (figure \ref{fig:SM}(a)) and accounting 
for the lower sensitivity of the MSSM $\h\Z$ searches, one can conclude 
that the same exclusion can be achieved with about 50\,$\pb$ per 
experiment if $\ss=192\,\G$.  Similarly, by running at $\ss=205\,\G$ with
120\,$\pb$ per experiment the four experiments could exclude an MSSM 
signal up to about $\mh = 110$\,GeV/$c^2$, thus eliminating a very large 
fraction of the lower $\tanB$ range allowed in the b--$\tau$ unification 
scenario.

Attention should be drawn to the fact that the above results were 
obtained for $\mt=175$\,GeV/$c^2$.  Varying $\mt$ by $\pm\,5$\,GeV/$c^2$ 
has the effect of shifting the theoretical upper bound on $\mh$ by about
$\pm\,5$\,GeV/$c^2$ at low $\tanB$ \cite{YellowBook}.

\vspace{0.4cm} 
Finally, we point out that the exclusion/discovery curves used here 
assume that a given luminosity was accumulated at some $\ss$ and do not 
take into account data gathered at lower energies.  As experimental 
searches usually retain some sensitivity even at energies above that at 
which they were originally applied, this would have the effect of slightly 
improving the SM and MSSM results presented here.  This can, in fact, be 
seen from figure \ref{fig:SM}(a), where the current 95\% C.L. exclusion 
limit on $\mHH$ after the collection of approximately 10\,$\pb$ of data by 
each experiment at both 161\,$\G$ and 172\,$\G$ exceeds the prediction for 
10\,$\pb$ collected at 175\,$\G$ alone.  It should also be noted that 
effort is continuously invested by all experiments in increasing their 
efficiencies for detecting a Higgs boson, which will result in further 
improvements in the mass sensitivity compared to these predictions.

\subsection*{References}
%


\section{Radiative Corrections to Chargino Production at \LEPTWO}

\author{M~A~Diaz, S~F~King, and D~A~Ross}

%
%
%
%
%

\subsection{Introduction}

The $e^+e^-$ colliders such as \LEP provide a clean environment for 
searching for the charginos predicted by the Minimal 
Supersymmetric Standard Model (MSSM) \cite{MSSMrep}. Several authors 
have considered at tree level the production of charginos 
at $e^+e^-$ colliders at the $Z$ pole \cite{prodLEP} and beyond 
\cite{prodLEPII}. On the other 
hand, from an accurate measurement of the chargino production 
cross-section, much information could be obtained about the MSSM 
\cite{others,us}. An intensive experimental search is being performed
with negative results, which translates into a lower bound on the 
chargino mass. According to the latest published results, we have 
taken $m_{\tilde\chi_1^{\pm}}>75$ GeV as long as the lightest neutralino
mass is not too close to the chargino mass \cite{expcha}.

Given the importance of a precise measurement of the chargino production
cross-section in $e^+e^-$ experiments, it is clearly necessary to
be able to calculate this cross-section as accurately as possible.
Although this cross-section does not contain any coloured particles,
and so is immune to QCD corrections, there are other radiative corrections
which, as we shall see, may give large corrections to the cross-section.
Although electroweak corrections may be expected to give contributions 
of order 1\%, there are additional radiative corrections coming from 
loops of top and bottom quarks and squarks which are important due to 
their large Yukawa couplings and it is 
these corrections which form the subject of the present paper. Radiative 
corrections to chargino masses have been calculated \cite{pierce}, 
nevertheless, the radiative corrections to chargino production in 
$e^+e^-$ experiments have not so been considered in the literature.

\subsection{The Calculation}

We consider the pair production of charginos with momenta $k_1$ and $k_2$
in electron-positron scattering with incoming momenta $p_1$ and $p_2$:
$$
e^+(p_2)+e^-(p_1)\rightarrow \tilde{\chi}^+_a(k_2) +\tilde{\chi}^-_b(k_1)
$$
In the MSSM charginos can be produced in the s--channel with intermediate
$Z$--bosons and photons, and in the t--channel with an intermediate 
electron-sneutrino $\tilde\nu_e$. 
The tree level cross section is determined by the
following parameters: the center of mass energy $\sqrt{s}$, the $SU(2)$ 
gaugino mass $M$, $\tan\beta$ defined as the ratio of the two Higgs vacuum 
expectation values, the supersymmetric Higgs mass $\mu$, and the 
sneutrino mass $m_{\tilde\nu_e}$. In practice we eliminate $|\mu|$ from
the set of independent parameters in favour of the lightest chargino mass
$m_{\tilde\chi_1^{\pm}}$.

\begin{figure}
\centerline{\protect\hbox{\psfig{file=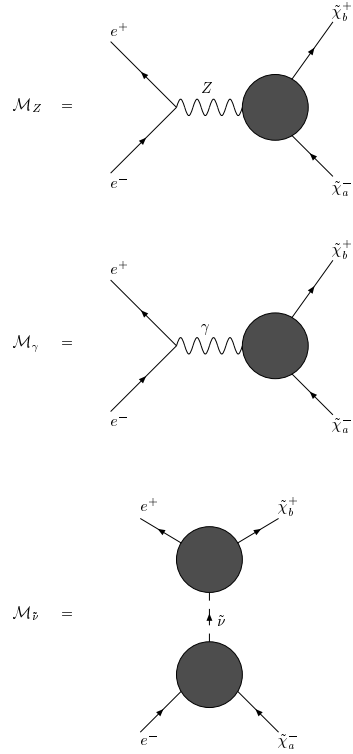,height=10cm,width=0.35\textwidth}}}
\caption{One--loop renormalized $M_Z$, $M_{\gamma}$ and 
$M_{\tilde\nu_e}$ amplitudes.} 
\label{ZGSneu1lAmplitudes} 
\end{figure} 
In our calculation of the one-loop radiative corrections
we work in the approximation where only top and bottom quarks and squarks
are considered in the loops. This implies, for example, that the 
electron--positron vertices, $\Gamma_{Zee}^{\mu}$ and 
$\Gamma_{\gamma ee}^{\mu}$, do not receive triangular corrections, and the 
tree level vertex can be identified with the one--loop renormalized vertex. 
In the presence of radiative corrections,
the amplitude for $e^+e^-\rightarrow \tilde{\chi}^+_b\tilde{\chi}^-_a$
may be expressed as the sum of three amplitudes $M_Z$, $M_{\gamma}$,  
$M_{\tilde{\nu}}$ as shown in Fig.~\ref{ZGSneu1lAmplitudes}. The shaded bubbles
in that figure are one--loop renormalized total vertex functions defined as
$i{\cal G}_{Z\chi\chi}^{ab}$, $i{\cal G}_{\gamma\chi\chi}^{ab}$ and
$i{\cal G}_{\tilde{\nu_e}e\chi}^{\pm a}$, respectively.
In the total vertex functions we include the tree level vertex, the
one--particle irreducible vertex diagrams plus the vertex counterterm, and
the one--particle reducible vertex diagrams plus their counterterms.
We work in the $\overline{MS}$ scheme, where the parameters in the
lagrangian are promoted to running parameters, working at the scale
$Q=m_Z$. Nevertheless, it should be 
stressed that the chargino mass values presented here correspond to the
pole mass. Although the detailed expressions for the
total vertex functions is quite complicated, by exploiting the possible
Lorentz structures of the diagrams it is possible to
express them in terms of just a few form factors.   

\begin{figure}
\centerline{\protect\hbox{\psfig{file=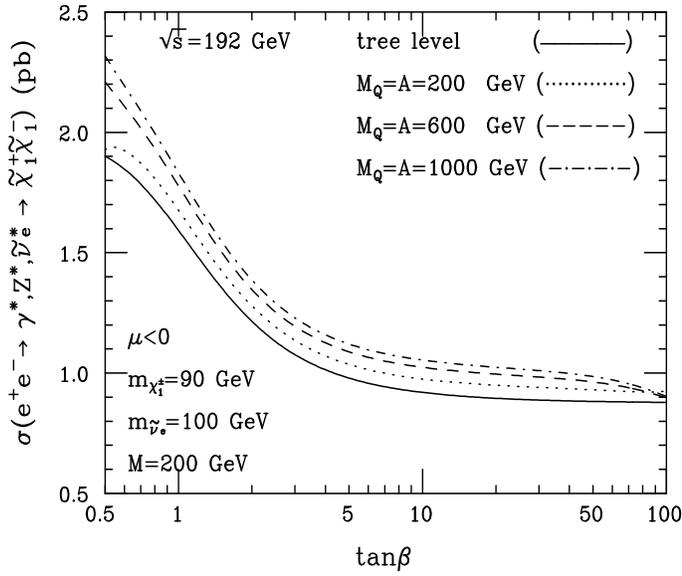,height=11cm,width=0.7\textwidth,angle=90}}}
\caption{One--loop and tree level chargino production cross section as a
function of $\tan\beta$.} 
\label{fig:c192tb} 
\end{figure} 

\subsection{The Results}

We present results for a center of mass energy of $\sqrt{s}=192$ GeV relevant
for \LEPTWO, and consider the case $\mu<0$. Radiative corrections to this 
cross section are parametrized by the squark soft masses which we take 
degenerate $M_Q=M_U=M_D$, and by the trilinear soft mass parameters 
$A\equiv A_U=A_D$, also taken degenerate. This choice is taken at the 
weak scale and it is made for simplicity, \ie, it should not be confused
with universality of minimal supergravity at the unification scale.

\begin{figure}
\centerline{\protect\hbox{\psfig{file=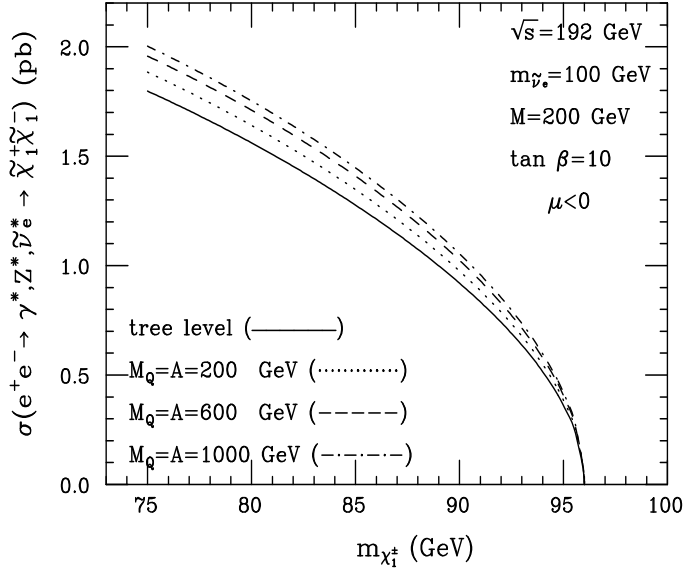,height=11cm,width=0.7\textwidth,angle=90}}}
\caption{One--loop and tree level chargino production cross section as a
function of the chargino mass $m_{\chi^{\pm}_1}$.} 
\label{fig:c192mc1}
\end{figure} 
In Fig.~\ref{fig:c192tb} we plot 
$\sigma (e^+e^-\rightarrow \tilde{\chi}^+_1 \tilde{\chi}^-_1)$ as a function
of $\tan\beta$, for a constant value of the chargino mass 
$m_{\chi^{\pm}_1}=90$ GeV, the electron-sneutrino 
mass $m_{\tilde\nu_e}=100$ GeV,
and the gaugino mass $M=200$ GeV. The tree level cross section is in
the solid line and decreases from 1.9 pb. to 0.9 pb. when $\tan\beta$ 
increases from 0.5 to 100. Three radiatively corrected curves are presented,
and they are parametrized by $M_Q=A=200$ GeV (dots), $M_Q=A=600$ GeV 
(dashes), and $M_Q=A=1$ TeV (dotdashes). For the chosen parameters we
observed that corrections are positive varying from a few percent at
large $\tan\beta$ to about 22\% at small $\tan\beta$. As expected, the
correction grows logarithmically with the squark masses, and are maximum 
when $M_Q=A=1$ TeV.

In Fig.~\ref{fig:c192mc1} we explore the dependence of the radiatively 
corrected cross section 
$\sigma (e^+e^-\rightarrow \tilde{\chi}^+_1 \tilde{\chi}^-_1)$ as 
a function of the chargino mass $m_{\chi^{\pm}_1}$. We fix the value
$\tan\beta=10$ and the rest of the independent parameters are taken as in 
the previous figure. Quantum corrections increase with the chargino
mass, starting from 11\% at $m_{\chi^{\pm}_1}=75$ GeV and increasing
slowly as we approach the kinematic limit for the chargino production,
where the total cross section drops to zero.

In conclusion, we have calculated radiative corrections to chargino pair 
production at \LEPTWO and found that one--loop contributions comming from 
top and bottom quarks and squarks increase the cross section by up to about 
20\%. Although a detailed exploration of the parameter space is being
carried out \cite{dkr}, the preliminary results presented here are enough to
conclude that quantum corrections must be included in order to extract
correctly the fundamental parameters of the theory from measurements
of the chargino mass and cross section \cite{us}.


\section{Charge and Colour Breaking Minima in the MSSM}
\author{ B~C~Allanach}

\begin{abstract}
Here, we review constraints that may be placed upon the parameters of the MSSM
by requiring that the vacuum lies in a bounded colour and charge conserving
global minimum of the scalar potential. The weakening of these constraints
originating from the possibility of a meta-stable charge and colour conserving
minimum are also presented.
\end {abstract}

\subsection{Introduction}

At a minimum of the scalar potential of a general model, the value of
each scalar field is either:
\begin{itemize}
\item{non-zero, in which case each symmetry under which the scalar field has
non-zero quantum numbers (i.e. the scalar is not a singlet) is spontaneously broken}
\item{zero, in which case the scalar does not correspond to any spontaneously
broken symmetries.}
\end{itemize}
The scalar potential of the MSSM includes, as well as the usual two Higgs
doublets that break the electroweak symmetry, squark and slepton fields.
At the minimum of the full scalar potential, if the vacuum
expectation values (VEVs) of charged sleptons are non-zero then
electromagnetism will be spontaneously
broken. If any squark fields have non zero VEVs, this corresponds to a
situation where QCD and electromagnetism have been broken~\cite{CCB}. This
obviously disagrees 
with a vast amount of experimental results which indicate that QCD and
electromagnetism are good symmetries and so this situation is empirically
ruled out. The terms that we will be interested in are:
\begin{eqnarray}
W &=& \mu H_1 H_2 + y_t Q_{L_3} H_2 t_R\ldots \nonumber \\
V_{soft} &=& \left( -y_t A_t \tilde{Q}_3 H_2 \tilde{t}_R + h.c.\right)
+  m_{\tilde{t}_L}^2 |\tilde{t}_L|^2 + m_{\tilde{t}_R}^2 |\tilde{t}_R|^2
+ m_2^2 |H_2|^2 + \nonumber \\ && \sum_{i=1,2,3} m_{\tilde{L}_i}^2
|\tilde{L}_i|^2
+ \ldots
\label{pot}
\end{eqnarray}
The scalar potential of squarks and sleptons is determined by the soft
parameters in the MSSM, and so certain ranges of these parameters have been
ruled out by several authors~\cite{CCB,sash1,callmun} on the grounds that they
would induce charge and
colour breaking (CCB) minima. A restrictive bound of this type involving top
squarks is
\begin{equation}
A_t^2+3 \mu^2 < 3 M^2,
\label{sim}
\end{equation}
where
\begin{equation}
M^2\equiv m_{\tilde{t}_L}^2+m_{\tilde{t}_R}^2.
\end{equation}

\subsection{Meta-Stable Vacuum}

There is however a possible get-out clause to some of these banned parameter
ranges. If the global minimum of the scalar potential is CCB but if there is a
local minimum of the scalar potential which conserves QCD and electromagnetism
but the global minimum is CCB, then we could not necessarily rule out the
scenario if the tunneling rate from the conserving to the CCB minima was
longer than the age of the universe. If this were the case, it is possible
that the universe rests in the meta-stable conserving vacuum and would never
tunnel through to the global minimum, thus being consistent with experiment.
The calculation of the tunneling rate has only been solved numerically by
calculating the 'bounce' solution~\cite{sash1}. The calculation is beyond the
scope of this
review, but it was noticed in ref.~\cite{sash1}, that roughly speaking, the
true
constraint from this effect changes the inequalities upon the MSSM parameters
to 
\begin{equation}
A_t^2+3 \mu^2 < 7.5 (m_{\tilde{t}_L}^2+m_{\tilde{t}_R}^2)
\label{tun}
\end{equation}
\begin{figure}
{\setlength{\epsfxsize}{0.4\vsize}
\setlength{\epsfysize}{0.4\vsize}
  \centerline{\epsfbox{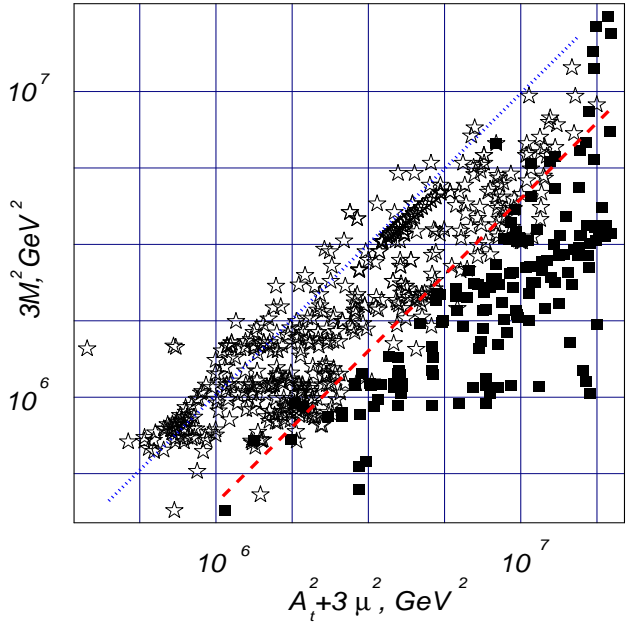}}}
\caption{The dotted line represents the simple criterion 
in Eq.\protect\ref{sim} for the absence of the
global CCB minima.  Taking into account the tunneling rates
relaxes this constraint to Eq.\protect\ref{tun} (roughly), shown as
the dashed line.  The scale is logarithmic.
}
\label{fig}
\end{figure}
It should be noted that if the relation in Eq.\ref{tun} is used, 
a few points in parameter space that should be banned will be allowed and
vice versa but on the whole Eq.\ref{tun} provides a good rule of thumb.
This is illustrated~\cite{sash1} in Fig.~\ref{fig}, where all of the
plotted points correspond to a CCB global minimum. The plot is really a
hyperplane through a multi-dimensional space of MSSM parameters. The stars
correspond to the case where there exists a meta-stable charge and colour
conserving vacuum whose lifetime is long compared to the age of the
universe. The boxes indicate points for which the conserving vacua should have
decayed into the CCB vacuum via quantum tunneling.

Aside from the constraints coming from the CCB minima, we can also place
bounds upon MSSM parameters which incur unphysical potentials that are
unbounded from below
(UFB)~\cite{callmun}. This occurs when the minimum of the potential lies
on a value of infinity for the VEV of a scalar field. The first (tree-level)
UFB constraints found~\cite{UFB} were 
\begin{equation}
m_2^2 - \mu^2 + m_{\tilde{L}_i}^2 \ge 0.
\end{equation}
In ref~\cite{callmun}, several CCB and UFB minima are found in addition to the
one
discussed above (including the other squarks etc.) 
One loop corrections are also discussed there.

\newpage

\section{Effect of R-Parity Violation upon Unification Predictions}
\author{B~C~Allanach and H~Dreiner}

\begin{abstract}
We present preliminary results on the possible effects of a particular,
possibly large, R-parity
violating (RPV) coupling in the MSSM. We focus on the effects upon
the unification predictions of $\tan \beta$ and $\alpha_S(M_Z)$, as compared
to the R-parity conserving (RPC) case.
We find that $\alpha_3 (M_Z)$ can be lowered by $\sim 3\%$, bringing the
prediction more in line with data. Bottom-tau Yukawa unification becomes
possible for any value of $\tan \beta$, rather than the restricted ranges
allowed in the RPC model.
\end{abstract}

\subsection{R-parity Violation}
In the RPV scenario, many of the RPV couplings are required to be very small
due to various constraints arising from flavour changing neutral
currents~\cite{bhatt}.
However, data still allows some of the couplings to be
large, of order 1. Here, we shall focus upon the effects of one RPV
interaction for which the strongest bound is often merely the perturbative
limit: 
\begin{equation}
W = \lambda_{333}' L_3 Q_3 D_3.  \label{superRPV}
\end{equation}
We consider the case where the full superpotential is that of the MSSM plus
the piece in Eq.\ref{superRPV}.
$\lambda_{333}'$ affects the renormalisation group evolution of the gauge
couplings to two loops and the third family Yukawa couplings of the MSSM to
one loop order. 

It is well known that grand unified theories (GUTs)~\cite{guts} predict 
\begin{equation}
\frac{3}{5} \alpha_1 (M_U) = \alpha_2 (M_U) = \alpha_3 (M_U),
\label{gaugeun}
\end{equation}
i.e. the unification of the gauge coupling
constants at the GUT scale $M_U$. We will view~\cite{lang} Eq.\ref{gaugeun} as
leading to
predictions of $\alpha_S(M_Z), M_U$ from the inputs $\alpha_{em}(M_Z)=127.9$
and $\sin^2 \theta_w=0.2315$.
\begin{figure}
\begin{center}
\leavevmode
\hbox{%
\epsfxsize=4.2in
\epsfysize=2.8in
\epsffile{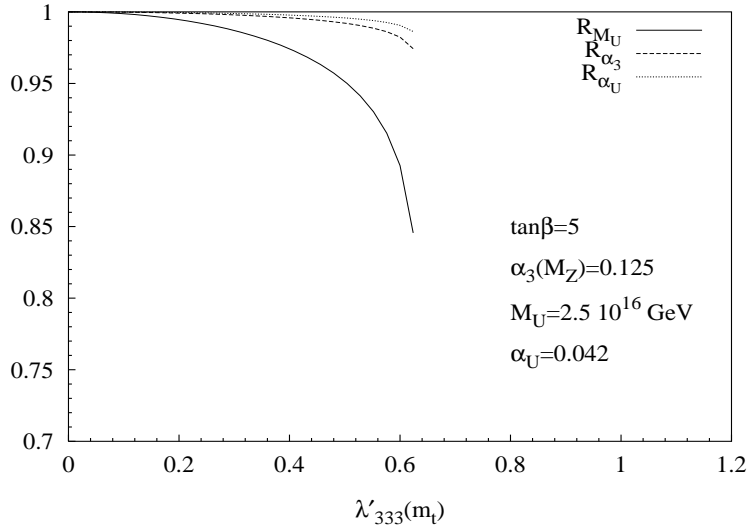}}
\end{center}
\caption{Effect of the RPV coupling $\lambda_{333}'$ upon unification
predictions. Input/output parameters are indicated on the figure.}
\label{fig:LD1}
\end{figure}
Using these inputs in the RPC MSSM we obtain (to two loop accuracy):
\begin{eqnarray}
\left(\alpha_3(M_Z)\right)^{RPC} &=& 0.125 \nonumber \\
\left(\alpha_i(M_U)\right)^{RPC} &=& 0.042 \nonumber \\
\left( M_U \right)^{RPC} &=& 2.46 \times 10^{16} \mbox{~GeV}. \label{MS}
\end{eqnarray}
Note that while the prediction of $\alpha_3(M_Z)$ in Eq.\ref{MS} seems to be
in disagreement with the present experimental value of $0.118 \pm 0.003$, much
of the discrepancy can be explained by
approximations such as the assumptions of a degenerate SUSY spectrum at
$m_t=176$ GeV and the absence of high scale threshold effects, which may be as
high as $+0.01$~\cite{thresh}.

GUTs also predict the relation
\begin{equation}
\lambda_b(M_U)=\lambda_\tau(M_U), \label{btau}
\end{equation}
where $\lambda_b,\lambda_\tau$ are the bottom and tau Yukawa
couplings respectively. Using inputs $m_b(m_b)=4.25 \pm 0.15$ GeV,
$m_\tau(m_\tau)=1.777$ GeV and $m_t$, we turn Eq.\ref{btau} into a prediction
for $\tan \beta$. In the RPC MSSM~\cite{guts}, there are two possible
ranges of $\tan
\beta$\footnote{$\tan
\beta$ is A SUSY
parameter of
the ratio of two Higgs vacuum expectation values.} consistent with this constraint: $1<\tan \beta<3$ or $\tan \beta>40$. 
\begin{figure}
\begin{center}
\leavevmode
\hbox{%
\epsfxsize=4.2in
\epsfysize=2.8in
\epsffile{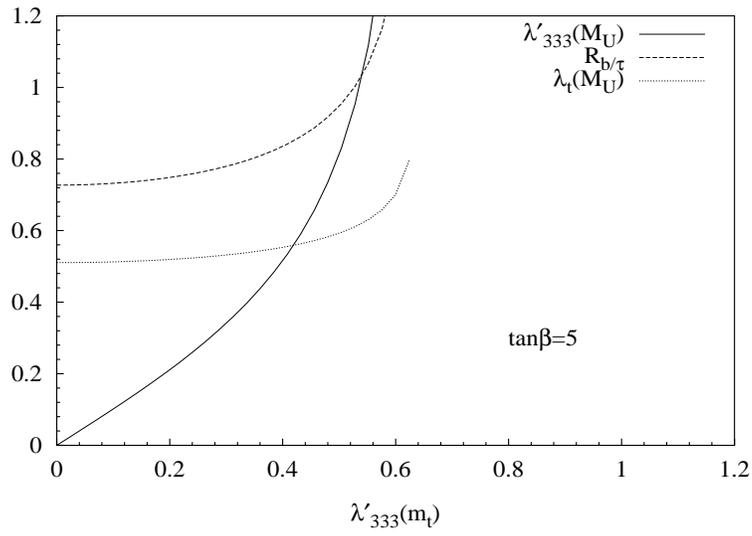}}
\end{center}
\caption{Dependence of GUT scale predictions upon $\lambda_{333}'(m_t)$}
\label{fig:LD2}
\end{figure}

To illustrate the difference between the RPC and RPV cases, we form the ratios
\begin{eqnarray}
R_{M_U} &\equiv& \frac{(M_U)^{RPV}}{(M_U)^{RPC}} \nonumber \\
R_{\alpha_U} &\equiv& \frac{(\alpha_{GUT})^{RPV}}{(\alpha_{GUT})^{RPC}} \nonumber \\
R_{\alpha_3} &\equiv& \frac{(\alpha_3(M_Z))^{RPV}}{(\alpha_3(M_Z))^{RPC}}
\nonumber \\
R_{b/ \tau} &\equiv& \frac{\lambda_b(M_U)}
{\lambda_\tau(M_U)}
\label{defns}
\end{eqnarray}
\begin{figure}
\begin{center}
\leavevmode
\hbox{%
\epsfxsize=4.2in
\epsfysize=2.8in
\epsffile{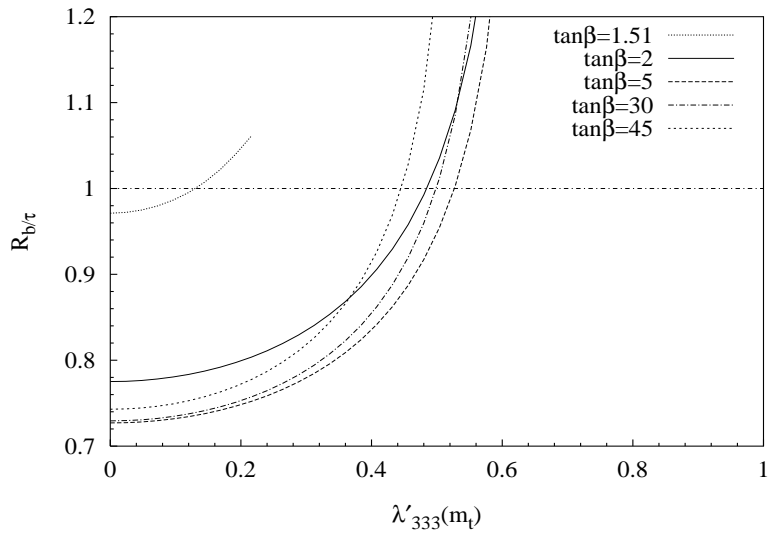}}
\end{center}
\caption{$R_{b/\tau}$ for various different values of $\tan \beta$. The
horizontal line at 1 marks the GUT prediction $R_{b/\tau}=1$.}
\label{fig:btau}
\end{figure}
where $\lambda_t$  is the top quark Yukawa coupling. 
We now pick $\tan \beta=5$, for which the RPC MSSM cannot unify the bottom and
tau Yukawa couplings while maintaining the correct top quark mass.
We choose various values of $\lambda_{333}'(m_t)$ and perform a semi-two loop
analysis\footnote{The running of gauge couplings and MSSM Yukawa couplings is
to two loops. $\lambda_{333}'$ is only run at one loop accuracy.
The results from a full two-loop calculation will be published
elsewhere~\cite{us2}.} to calculate the $R_i$ defined in Eq.\ref{defns} from
the constraints in Eqs.\ref{gaugeun},\ref{btau}.
Fig.\ref{fig:LD1} shows 
that $\alpha_S(M_Z)$ can be decreased by up to 3\% by adding the RPV coupling,
the unification scale by
15\% and the unified gauge coupling by 2\%.
The perturbative limit is shown as the right-hand end of the lines in 
Fig.\ref{fig:LD2}, where $\lambda_{333}' \approx 0.6$. $\lambda_t(M_U)$ is
shown to be large, but has to be smaller than the case of RPC to yield the
same top quark mass.
Fig.\ref{fig:LD2} also shows that for $\tan \beta=5$ and $\lambda_{333}'(m_t) =
0.53$, it is possible to satisfy the bottom-tau Yukawa unification constraint,
unlike in the RPC MSSM. In fact, Fig.\ref{fig:btau} illustrates that for {\em
any} value of $\tan \beta$, a value of $\lambda_{333}'(m_t)$ may be chosen to
yield $R_{b/\tau}=1$.


\section{Searches for Pair-Produced New Particles: Trade-Off Between
  Integrated Luminosity and Centre-Of-Mass Energy }
\author{T~Wyatt}

\begin{abstract}
When searching for massive, pair-produced, new particles close to the
kinematic limit at \LEP, achieving the maximum possible centre-of-mass energy
$\Ecm$  has the benefit of giving the highest possible 
production cross-section.
However, because of, e.g., operational instabilities when running at
the highest possible energy, a larger integrated luminosity ($\IL$) may be
achieved when running at a somewhat lower $\Ecm$.
It therefore becomes interesting to consider the possible trade-off
between $\Ecm$ and $\IL$ .
\end{abstract}

\subsection{Introduction}
It is impossible to predict with any certainty the actual choices that
we may be faced with in 1999/2000.
For example, how much luminosity will we already have collected by
then
and at what centre-of-mass energy?
How stable will the operation of the superconducting cavity system
have become?
Also, the dependence of cross-section on $\Ecm$ 
depends on the spin and mass of 
the pair-produced, new particles being considered.
The selection efficiency and expected background level also depends on
these and possibly other factors (such as,
in searches for supersymmetric particles,
 the mass of the Lightest Supersymmetric Particle).
However, by considering a few specific cases it is possible to notice
some interesting features and to draw some general conclusions
that should be largely independent of these details.

\subsection{Examples}
\begin{figure}[htbp]
 \epsfxsize=8cm
\begin{center}
\mbox{\epsffile{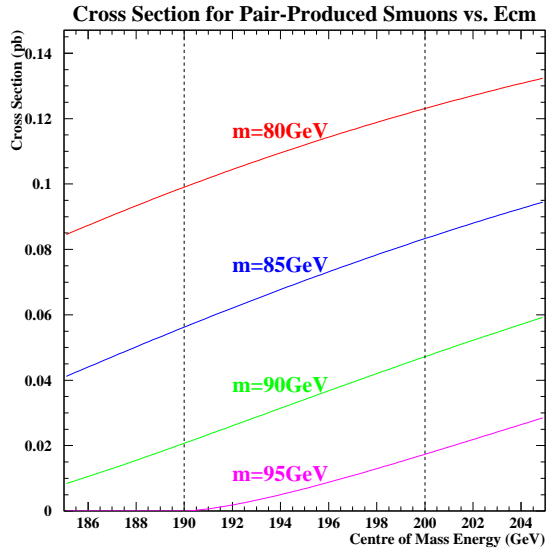}}
\end{center}
 \caption{
The expected production cross-section for smuon pairs as a function of
$\Ecm$ for various smuon masses, assuming a $\beta^3/s$ dependence of
the cross-section.
} 
\label{fig-slept}
\end{figure}

As a first example, let us consider the production of smuon pair
events.
\begin{center}
$\ee \rightarrow \tilde{\mu}^+ \tilde{\mu}^-$,

$\tilde{\mu}^\pm \rightarrow  {\mu^\pm} \nt_1$,
\end{center}
The standard experimental signature  is the observation of events
containing a pair of muons and significant missing momentum transverse
to the beam direction.
The dominant backgrounds arise from two-photon processes and W pairs
and are roughly independent of $\Ecm$.
The expected selection efficiency depends on \msmu\ and \mchz, but for
given values of these parameters is roughly  independent of $\Ecm$ .
The sensitivity of the search therefore depends mainly on  $\IL$ and
the expected production cross-section, which is approximately
proportional to  $\beta^3/s$ (shown in figure~\ref{fig-slept}).

If we take as an example \msmu\ =~90~GeV we can see that the expected
cross section at $\Ecm$ =~200~GeV is roughly double that at $\Ecm =~190~$GeV.
This might lead us naively to conclude that in searching for smuons with
$\msmu\ =~90~$GeV we could afford to loose only a factor of two in \IL\
in exchange for the higher energy.
In fact this would be incorrect, because it neglects the fact that we
will already have collected an appreciable amount of luminosity at 
$ \Ecm =~190~$GeV.
As we will see below, this increases the relative value of new data at
a higher $\Ecm$ compared to merely increasing the integrated luminosity
already collected at a lower energy.

We will calculate the expected 95\% CL lower limit on \msmu\
obtainable by combining the data from the four \LEP experiments under
the following assumptions:
\begin{enumerate}
\item
The expected signal efficiency is 50\% and the remaining background
from \smp\ is 0.1~pb independent of $\Ecm$.
(This corresponds to a reasonable guess for the performance in the
region of low \mchz, where the signal events are rather difficult to
distinguish kinematically from the W pair background.
\item
An integrated luminosity of $200\ipb$ per experiment
has already been collected at 
$\Ecm =~190~$GeV and the limit is calculated by combining this data with
new data at three possible values of $\Ecm : 190, 195$ or $200~$GeV.
\item
The number of observed candidate events is equal to the number
expected due to background
from \smp.
Background is subtracted using the standard PDG recipe.
\end{enumerate}

\begin{figure}[htbp]
 \epsfxsize=8cm 
\begin{center}
\mbox{\epsffile{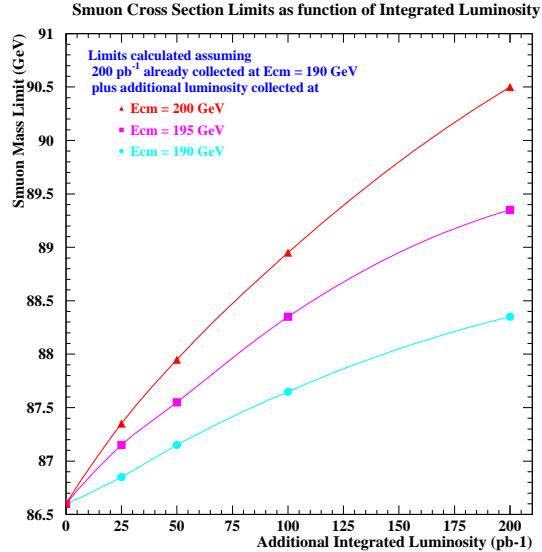}}
\end{center}
 \caption{
The expected 95\% CL lower limit on \msmu\
obtainable by combining the data from the four \LEP experiments,
plotted as a function of
the {\em additional} integrated luminosity collected per experiment.  
The three curves correspond to three possible choices of $\Ecm$ at
which this  {\em additional} luminosity is collected.
} 
\label{fig-slimit}
\end{figure}

The results are given in  figure~\ref{fig-slimit}.
The expected 95\% CL lower limit on \msmu\ is plotted as a function of
the {\em additional} integrated luminosity collected per experiment.  
The three curves correspond to three possible choices of $\Ecm$ at
which this  {\em additional} luminosity is collected.
By comparing the curves for  $\Ecm =~190~$GeV and $\Ecm =~200~$GeV we can
see that if the additional data is collected at the
higher  $\Ecm$ we can achieve the same limit on \msmu\ with only about
one third as much integrated luminosity.
The signal:noise ratio is larger in the higher energy data and this is
why it wins out by a larger factor than the simple ratio of the
expected production cross-sections.

As a second example, let us consider the pair production of charginos.
For this search the arguments in favour of higher $\Ecm$ over higher \IL\
are overwhelming.
Charginos are fermions and, therefore, are produced with a cross-section
proportional to $\beta/s$, which has a very sharp turn-on above the
kinematic threshold.
The expected production cross-section is model dependent, but there is
a sizable cross-section (of the order of 1~pb) over much of
the parameter space of the MSSM.
For example,
with only 10~\ipb\ collelcted at  $\Ecm =~172~$GeV the \LEP experiments
have already been able to place stringent limits on chargino
production at close to the kinematic limit.
(See the talk of Sylvie Rosier-Lees at the XXXII$^{\mathrm{nd}}$ Rencontres
de Moriond, Les Arcs, France, March 1997.)
Of course, one can always pick a region of parameter space which
results in very small production cross-section or detection
efficiency, but it seems more likely that we will discover charginos by
collecting a few tens of \ipb\ at a
new high $\Ecm$  than by further running at an $\Ecm$ at 
which one already has collected hundreds of \ipb . 

In conclusion, when searching for particles that do not show a sharp
rise of the production cross-section at the kinematic threshold, the
potential loss in integrated luminosity that might result from running
at the highest possible $\Ecm$ must be taken into account when
assessing the optimal running strategy.
When performing such studies it is important to take into account the
fact that we will already have collected (hopefully) a few hundred
\ipb\ at around  $\Ecm =~190~$GeV and that the final limit/discovery 
will result from the combination of all the available data.
Failure to do this may underestimate the relative worth of collecting
data at the highest possible $\Ecm$ .
In searching for particles such as charginos, which may well be produced with
a large cross-section that rises sharply  at the kinematic threshold
one would have to bet on the highest possible $\Ecm$ as the factor most
likely to produce a discovery.


\section{Single Sneutrino Production at \LEPTWO} 
\author{ B.C. Allanach, H. Dreiner, P. Morawitz and M.D. Williams}

\begin{abstract}
We propose a new method of detecting supersymmetry at \LEPTWO when R-parity is 
violated by an $LLE$ operator. We present the matrix element for the process 
$\gamma e \rightarrow \tilde{\nu}_j e_k$ and calculate the cross-section in
$e^+e^-$ collisions. 
A preliminary Monte-Carlo analysis is undertaken to investigate the possible
sensitivity to this signal and we present the
5$\sigma$ discovery contours in the $m_{\tilde{\nu}_j}$ vs.\ coupling plane.
\end{abstract}

\subsection{Introduction}

If R-parity is violated it becomes possible for supersymmetric particles
to be produced singly via the $\rpv$ couplings. Although the production
cross-section is suppressed by the coupling, the higher kinematic reach
means that this can provide a complimentary approach to the study of
pair production. Here, we consider the possible detection of R-parity 
violating supersymmetry through the operator\footnote{Here $L$ and $E$ are the
SU(2) doublet and singlet lepton superfields respectively.} 
\begin{equation}
W_{LLE} = \lambda_{ijk} L_i L_j E^c_k \label{LLEsup}
\end{equation}
where $i,j,k$ are family indices and gauge indices have been suppressed.
This operator has been studied previously \cite{direct}
in the context of resonant 
production of single sneutrinos via the couplings $\lambda_{121}$ and 
$\lambda_{131}$ and in terms of alterations to the distributions of 
dilepton events. The mechanism that we propose is not as effective for
these couplings, but is applicable to other couplings. The scenario we 
consider is one in which a photon from one of the beam electrons interacts
with the other beam electron to produce a single sneutrino and a lepton. 

\subsection{The Matrix Element}

\begin{figure}
\begin{center}
\begin{picture}(300,100)(0,0)
\put(0,0){\begin{picture}(150,100)(0,0)
\Photon(0,100)(50,50){3}{5}
\ArrowLine(0,0)(50,50)
\ArrowLine(50,50)(100,50)
\ArrowLine(100,50)(150,100)
\DashLine(100,50)(150,0){5}
\Text(75,48)[tc]{$e_j$}
\Text(25,23)[tl]{$e_1(p_2)$}
\Text(25,77)[bl]{$\gamma(p_1)$}
\Text(125,23)[tr]{$\tilde{\nu}_k(q_2)$}
\Text(125,77)[br]{$e_j(q_1)$}
\end{picture}}
\put(200,0){\begin{picture}(100,100)(0,0)
\Photon(17,100)(50,66){3}{5}
\ArrowLine(17,0)(50,33)
\ArrowLine(50,33)(50,66)
\ArrowLine(50,66)(83,100)
\DashLine(50,33)(83,0){5}
\Text(52,50)[cl]{$e_j$}
\Text(34,19)[br]{$e_1(p_2)$}
\Text(34,81)[tr]{$\gamma(p_1)$}
\Text(67,19)[bl]{$\tilde{\nu}_k(q_2)$}
\Text(67,81)[tl]{$e_j(q_1)$}
\end{picture}}
\end{picture}
\end{center}
\caption{Contributing diagrams. The other initial lepton has been neglected.}
\label{fig:verts}
\end{figure}
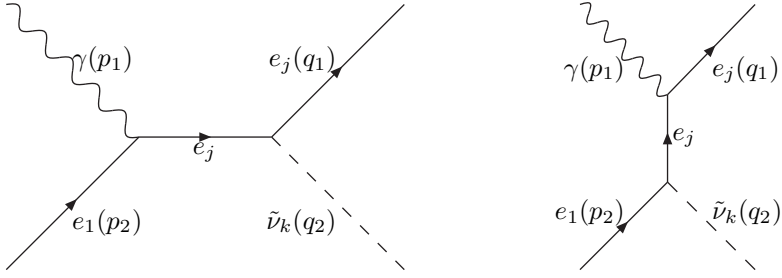

Using the Weisz\"{a}cker-Williams approximation~\cite{WW}, 
the appropriate diagrams are shown in Figure~\ref{fig:verts}.
We now calculate the matrix element for the process
\begin{eqnarray}
\gamma(p_1)+ e_1(p_2)  &\rightarrow&  e_y(q_1) +\tilde{\nu}_x(q_2) \nonumber \\
e_y=\mu,\tau & & \tilde{\nu}_x=\tilde{\nu}_{\mu,\tau}.
\end{eqnarray}
Neglecting fermion masses, the Mandelstam variables are then defined as
\begin{eqnarray}
\begin{tabular}{ccc}
$s = (p_1+p_2)^2$ & $t = (p_1-q_1)^2$ &$ u = (p_1-q_2)^2$  \\
\end{tabular}
\end{eqnarray}
where we have assumed the Weisz\"{a}cker-Williams approximation of an on-shell
photon and a real sneutrino.
The matrix element squared (where the photon is approximately on-shell) is
\begin{equation}
|\bar{M}|^2 = e^2 \lambda^2 \left[1 + \frac{u}{t} + \frac{u}{s} +
\frac{u^2}{st} - \frac{t}{2 s} - \frac{s}{2t} \right]. \label{LLEmat}
\end{equation}
$\lambda$ is the R-parity violating coupling probed. The matrix
element in Eq.\ref{LLEmat} is spin-averaged.

\subsection{Cross-section Evaluation}

The cross-section for $e^+e^- \rightarrow el\tilde{\nu}$ is obtained from that
for $\gamma e \rightarrow l \tilde{\nu}$ by 
\begin{equation}
\sigma(s;e^+e^- \rightarrow el\tilde{\nu}) = \int f_\gamma(y) \sigma(ys;\gamma e \rightarrow l \tilde{\nu}) dy
\end{equation}
where $f_\gamma(y)$ is the photon distribution in the electron at a given 
fraction, $y$, of the electron momentum. We use the following version of the
Weizs\"{a}cker-Williams distribution \cite{Frixione}: 
\begin{eqnarray}
f_\gamma(y) & = & \frac{\alpha_{em}}{2\pi} \left\{ 2(1-y)
                  \left[\frac{m_e^2 y}{E^2(1-y)^2\theta_c^2 + m_e^2y^2} - \frac{1}{y} \right] \right. \nonumber \\
            &   & \left. + \frac{1+(1-y)^2}{y} \log\frac{E^2(1-y)^2\theta_c^2 + m_e^2y^2}{m_e^2y^2} \right\}
\end{eqnarray}
where $\theta_c$ is the maximum scattering angle of the beam electron and $E$ is
the beam energy. We take the value of $\theta_c$ to be $30 \, {\rm mrad}$ which is 
a typical value for the coverage of the luminosity monitors in a \LEP experiment.
If we were to allow the full range of scattering angles the cross-sections would be 
higher by about $20\%$, but the Weizs\"{a}cker-Williams approximation would not be
such a good one.

\begin{figure}
\centering 
\epsfig{file=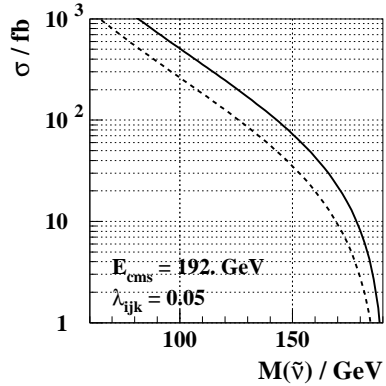,width=5.5cm}
\caption{The cross-section for single sneutrino production at a centre of mass energy
of $192 \, {\rm GeV}$ and for $\lambda_{ijk} = 0.05$ as a function of the sneutrino
mass. The solid line is the cross-section when the final state lepton is a muon;
the dashed line is the cross-section when the final state lepton is a tau.}
\label{single.xsect} 
\end{figure}

If fermion masses are neglected the differential cross-section has a singularity 
at $t=0$, but this can be avoided by including the mass of the exchanged fermion
in the divergent diagrams. This results in a larger total cross-section for the
process $e^+e^- \rightarrow e \mu \tilde{\nu}$ than for 
$e^+e^- \rightarrow e \tau \tilde{\nu}$. The two cross-sections are plotted as 
functions of the sneutrino mass in Fig.\ref{single.xsect} for a value of the
coupling $\lambda_{ijk} = 0.05$. 

\subsection{Final States from Single Sneutrino Production}

The signal $e^+e^- \rightarrow e l \tilde{\nu}$ is characterised by the electron
continuing along the beam pipe, so that the only particles visible in the detector
are the lepton $l$ and the decay products of the sneutrino. The sneutrino can either
decay directly, $\tilde{\nu} \rightarrow e l$, or indirectly via lighter
charginos or neutralinos, eg. $\tilde{\nu} \rightarrow \nu \chi$. The final state
depends upon both the coupling involved and the flavour of sneutrino produced. This
information is summarised in Table \ref{finalstate}.

\begin{table}
\centering
\begin{tabular}{|c|c|c|c|c|}
\hline
Coupling & \multicolumn{2}{c|}{Direct Decays} & \multicolumn{2}{c|}{Indirect Decays} \\ \cline{2-5}  
         & $\tilde{\nu}_{\mu}$ & $\tilde{\nu}_{\tau}$ & $\tilde{\nu}_{\mu}$ & $\tilde{\nu}_{\tau}$ \\ \hline
 122     &    $e \mu^+\mu^-$   &           -          &  $\mu \nu  \chi$    &          -           \\
 123     &   $e \tau^+ \tau^-$ &           -          &  $\tau \nu  \chi$   &          -           \\
 132     &          -          &    $e \mu^+\mu^-$    &         -           &  $\mu \nu  \chi$     \\
 133     &          -          &   $e \tau^+ \tau^-$  &         -           &    $\tau \nu  \chi$  \\
 231     &  $e \tau^+ \tau^-$  &    $e \mu^+\mu^-$    &  $\tau \nu  \chi$   &  $\mu \nu  \chi$     \\ \hline
\end{tabular}
\caption{The final states for production of different sneutrino flavours via different couplings.
The entries marked with a dash are those for which that sneutrino flavour cannot be produced by
that coupling.}
\label{finalstate}
\end{table}

For a coupling $\lambda_{ijk}$ the lightest neutralino can decay to the following 
final states:
\begin{displaymath}
\tilde{\chi}^0_1 \rightarrow \left\{ 
\begin{array}{c}
l_i^- \bar{\nu}_j l_k^+ \\
l_i^+     {\nu}_j l_k^- \\
\bar{\nu}_i l_j^- l_k^+ \\
    {\nu}_i l_j^+ l_k^- 
\end{array} \right.
\end{displaymath}
So that the indirect decays via the lightest neutralino will contain three charged
leptons and two neutrinos.

\subsection{Investigation of Final State Signals}

To investigate the viability of searching for these signals we have written a Monte 
Carlo capable of generating the different final states and including interfaces to 
JETSET, for the decays, and PHOTOS, for final state radiation. Several simple analyses 
to discriminate the signal from the dominant backgrounds were developed.

\begin{figure}
\centering
$
\begin{array}{cc}
\epsfig{file=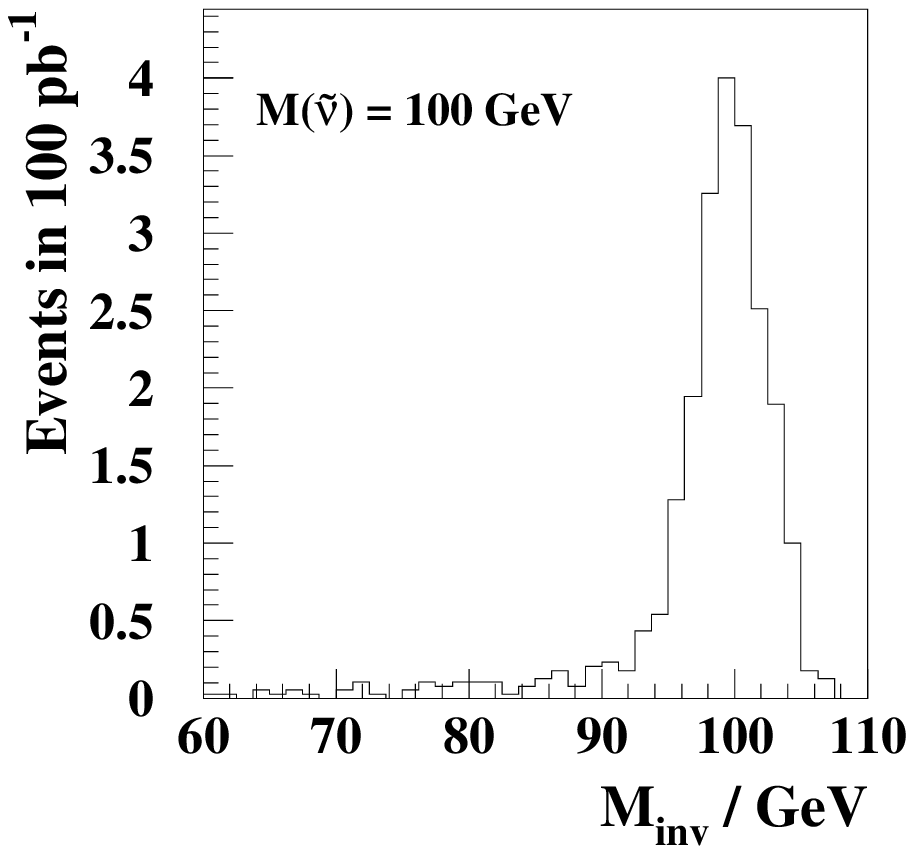,width=5.5cm} & \epsfig{file=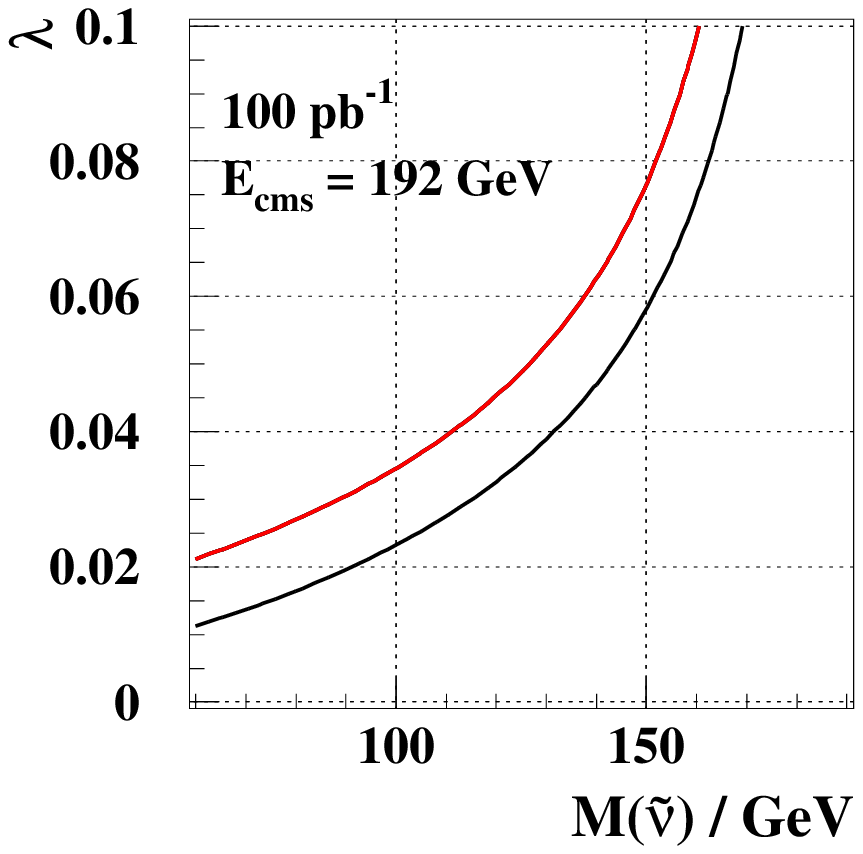,width=5.5cm} 
\end{array}
$
\caption{The left hand plot shows the distribution of the invariant mass of the 
electron and muon from a sneutrino decay. In the right hand plot the 
discovery coutours for the best case (indirect decays with $\lambda_{122}$)
are shown in black and for the worst case (direct decays to $e\tau$) in grey.}
\label{discovery}
\end{figure}

The most important backgrounds are two photon processes, but four fermion events are
also significant. Direct decays have a similar signature to 
$\gamma\gamma \rightarrow \mu\mu$ or $\tau\tau$ with a single tagged electron. In the 
case of the signal the electron is produced from the decay of a massive sneutrino
and is apparently scattered through a large angle. In addition, the average tranverse
momentum of the tracks is larger than for a typical $\gamma\gamma$ event. For direct
decays to $e\mu$ a large visible mass can be required and the invariant mass of the
sneutrino can easily be constructed.
The distribution of this invariant mass for a sneutrino of $100 \, {\rm GeV}$
is shown in Fig.\ref{discovery} where we have assumed an invariant mass resolution
of $2.5 \, {\rm GeV}$.
Large missing transverse momentum can be required for direct decays to $e\tau$.

For the indirect decays substantial missing energy 
is expected because of the presence of an energetic neutrino. This neutrino also
means that the missing momentum is often not along the beam pipe. The decay products of
the neutralino depend upon the choice of coupling. 
For a $\lambda_{122}$ coupling there are three leptons in the event and this can be
used to reduce the background. For a $\lambda_{133}$ coupling the visible mass and
total charged energy is small.
For a sneutrino of $100 \, {\rm GeV}$ and, for the cascades, a neutralino of $50\, {\rm GeV}$ the 
following performances are obtained by our selections:
\begin{itemize}
\item Direct decays to $e \mu$: efficiency of $45\%$ and background of $18 \, {\rm fb}$.
\item Direct decays to $e \tau$: efficiency of $43\%$ and background of $28 \, {\rm fb}$.
\item Indirect decays $\lambda_{122}$: efficiency of $49\%$ and background
of $8 \, {\rm fb}$.
\item Indirect decays $\lambda_{133}$: efficiency of $42\%$ and background of $23 \, {\rm fb}$.
\end{itemize}

Using these simple selections and parameterising the variation of the efficiency
with sneutrino mass we can derive expected discovery ($5\sigma$) contours in the plane 
$(m_{\tilde{\nu}}, \lambda)$. In Fig.\ref{discovery} these contours are shown assuming
$100 \, {\rm pb^{-1}}$ of data are collected at a centre of mass energy of $192 \, {\rm GeV}$.
A combination of the four \LEP experiments would significantly improve the results.
In addition, much better performance could be obtained for the case of direct decays
to $e\mu$ by including the mass distribution of the events.
These results should be compared with the available limits upon R-parity violating
couplings. The best current limits on $\lambda_{ijk}$ in Eq.\ref{LLEsup} 
are\footnote{The bounds on $\lambda_{12n}$ are at the $2\sigma$ level; 
all others are at $1\sigma$.}~\cite{bhatt2}
\begin{equation}
\begin{array}{ccccc}
\lambda_{12n} < 0.05 & \lambda_{131} < 0.06 & \lambda_{132} < 0.06 & \lambda_{133} < 0.004 & \lambda_{23n} < 0.06 
\end{array}
\end{equation}
for a SUSY mass of 100 GeV. It can be seen that these limits are not sufficient to
exclude the possibility that sneutrinos with masses beyond the kinematic limit for
pair production may be probed via this process.


\subsection{Conclusion}

We have calculated the matrix element for $\gamma e \rightarrow \tilde{\nu}_j e_k$
via an R-parity violating coupling of type $LLE$ and obtained the cross-section
in $e^+e^-$ collisions. The expected final states from such processes at 
\LEPTWO have 
been listed and a preliminary investigation made. In view of the encouraging 
results derived here, a future experimental analysis to address this possibility is
very welcome.

\subsection{Acknowledgements}

We are grateful for useful discussions with M. Seymour and M. Kraemer from which
this paper has benefitted.


\section{Searches at \LEPTWO  related to possible \HERA effect}
\label{sub1}
\author{G~W~Wilson}

\begin{abstract}
The recent observations by the \HERA 
experiments 
have revived interest in the phenomenology
of
particles with 
leptoquark like couplings. 
Leptoquarks or 
R-parity violating squarks in supersymmetric theories
or contact interactions 
were discussed as possible exotic explanations
of the observations, 
and we briefly review 
the relevance of
related indirect and direct searches at \LEPTWO. 
\end{abstract}

\subsection{Introduction}

This article addresses the following two questions :

\begin{enumerate}
\item{
{\bf IF} the excess of events at \HERA at high (x,Q$^2$)
in e$^+$ p $\rightarrow$ e$^+$ jet X \cite{H1,ZEUS} is confirmed
and establishes a particle of mass exceeding 180 GeV
{\bf THEN} is there a role for experiments at \LEPTWO 
in establishing or eliminating
some of the possibilities?}

\item{Independent of the actual relevance of possible \LEPTWO 
searches to the \HERA effect, are there some searches 
inspired by models discussed in this context that should be 
considered 
anyway at \LEPTWO  and have so far received little 
experimental attention?
This is most important if the experimental 
signature is not covered by the established search analyses. }
\end{enumerate}

There has been ``much ado about leptoquarks'' 
in recent preprints\cite{Hewett,preprints}.
Among the more relevant ones to \LEPTWO  
research are \cite{Hewett,Ellis,relevant}.
Perhaps the most compelling recent works 
include one\cite{Drees} which contests the mutual
consistency of the two \HERA experiments, let alone the 
consistency with a resonance, and a publication 
prior to \cite{H1,ZEUS} which suggests searching 
for single leptoquark production at \LEPTWO  in electron-photon
scattering\cite{Doncheski}.

\subsection{Related searches at \LEPTWO}
Many things can be done at \LEPTWO, but the reader should 
be careful to assess, whether such searches are complementary 
or not to similar searches at the Tevatron or at \HERA which 
have the advantage of a higher possible kinematic reach. 
The ``standard'' pair-production search possibilities
and estimated sensitivities
are discussed in \cite{YR}.

Leptoquark like new particles could be 
pair-produced at \LEPTWO  through their coupling to the 
photon and the Z. This has the great advantage of 
a production cross-section which is independent of the Yukawa 
coupling, and sensitivity to
all generations. (All \HERA limits on leptoquarks depend on this
coupling and are relevant to first generation leptoquarks).
But of course the mass reach is limited to the 
beam energy. So, one could imagine testing for all 
sorts of leptoquarks up to at most 100 GeV. 
This search faces stiff competition from the Tevatron
which also has the advantage of a cross-section 
independent of the Yukawa coupling.
For example the first generation sclar leptoquark limit 
for 100\% branching fraction to electron-quark is 225 GeV at 95\% CL \cite{D0}.
The weakest limit from the Tevatron in the $\ell$-$\ell$-jet-jet searches
is for a third 
generation leptoquark, 
the $\tau \tau$ jet-jet mode, (99 GeV from CDF \cite{CDF1}) - this 
would be worth doing better.
The presently only partially covered region by the Tevatron, 
partly
because it is more 
difficult experimentally, is
the search in the $\nu \overline{\nu}$-jet-jet mode 
for all three generations. Some leptoquark like objects,
have a 100\% branching ratio to neutrino-quark, because they
couple only to neutrinos and quarks, so 
making them 
inaccessible in single production from $\mathrm{ep}$ and $\mathrm{e}\gamma$ 
interactions
at \HERA and at \LEPTWO respectively.
The perfect place to look for these up to the beam energy is in pair 
production
at \LEPTWO.
This search is 
already well covered since the
experimental signature is identical to the Higgs missing 
energy channel.

As already discussed, the most intriguing prospect at \LEPTWO  for 
direct searches possibly relevant to the \HERA observation, is the 
single production of leptoquark like particles via electron-photon
scattering. This depends on the flux 
of Weizsacker-Williams photons from one beam (QED), the parton distribution 
function in the photon,
which parametrises the splitting probability and energy fraction
for the photon splitting to a quark-antiquark pair, and again $\lambda$, the
strength of the Yukawa coupling between the leptoquark-e-quark.
This method can test all the possible e-quark interactions 
at \HERA for both e$^+$ and e$^-$ and for charge 1/3 and 2/3 quarks 
and antiquarks. 
\OPAL\ has carried out a ``demonstrator'' preliminary analysis\cite{SSR}
using 20 pb$^{-1}$ at 161 and 172 GeV
with this
technique in both the electron-jet and neutrino-jet topologies (these
are complementary to existing searches) and find a limit of
131 GeV for scalar first generation leptoquarks of
charge 1/3 or 5/3
for couplings $\lambda > \sqrt{4 \pi \alpha_{\mathrm{em}}}$. 
It seems feasible to eventually probe Yukawa couplings of 
electro-magnetic strength up to about 10-20 GeV from the 
kinematic limit. This seems to be one more case amongst the many
good ones for pushing \LEPTWO  as far as possible in energy.
 
The most discussed method for
\LEPTWO  to constrain 
possible new physics beyond the \LEPTWO  kinematic limit
is from deviations in the measured cross-sections and angular 
distributions for two-fermion production. 
This field is usually considered in two approximations.
For new physics with mass scale $m_X \gg \sqrt{s}$ one 
parametrises the new physics as an effective four-fermion
contact interaction. This approach has been followed 
for many years. The most recent experimental publication 
relevant to this is \cite{OPAL_161}.
For cases where the mass scale is not much higher than 
the centre-of-mass energy, one needs to take into account 
the diagram from t-channel exchange of 
a virtual leptoquark like particle in the 
process e$^+$e$^-$$\rightarrow$q$\overline{\rm{q}}$. See \cite{Hewett} 
for more details.
All \LEP experiments are fairly active in pursuing this
line, and the 
finalised publications containing the multi-hadron cross-section
measurement 
from 1996 \LEPTWO  running are likely to 
address both possibilities.
This approach looks to be quite constraining on the possible 
interpretation of the \HERA observation arising from e$^+$/sea-quark
interactions with large $\lambda$ \cite{Ellis}.
However the much lower values of $\lambda$,
$\lambda \approx 0.1 \sqrt{4 \pi \alpha_{\mathrm{em}}}$, 
implied by considering models with e$^+$/valence-quark at \HERA
will not be constrained by future \LEPTWO data with
this approach.

\subsection{Summary}

There are quite a few topics which
can be addressed at \LEPTWO  in the context of possible 
new physics being accessed at \HERA.
There are also many searches  
awaiting the eager searcher keen to discover/constrain
R-parity violating supersymmetry \cite{HERE}.
However, it is difficult to conceive of an experimental
topology which could not be produced by
R-parity violating supersymmetry.

It is most important to repeat the 
\HERA experiments and establish if anything is actually awry.
However, the obvious conclusion is that direct searches at \LEPTWO 
with presently foreseen beam energies
have no role if the only new particles 
that exist have mass above about 200 GeV.
In that case, a higher energy e$^+$e$^-$ collider
would be an ideal tool to explore such phenomena\cite{Hewett}.

The main new experimental signatures,
suggested by single leptoquark production,
are the lepton-jet and neutrino-jet topologies.


\section{Summary}
\author{ J~Ellis}

\begin{abstract}
Aspects of searches at \LEPTWO are reviewed and summarized, with
particular emphasis on the gains from running \LEP at 200 GeV, and on
alternative paradigms for supersymmetric phenomenology, such as
models with violation of $R$ parity.
\end{abstract}

\subsection{Introduction}

The primary search at \LEPTWO~\cite{LEP2YB} is that for the Higgs boson of the 
Standard
Model~\cite{ZH}, whose prospects are discussed in section 2 of this report. 
Many other new particles have been conjectured, and may also be
hunted at \LEPTWO under exceptionally clean and well-understood conditions.
During the time available, this Working Group was unable to study all
these alternatives, and made a selection of quarries that was
guided largely by the subjective research interests of the participants.

Foremost among these interests was supersymmetry~\cite{susy}, which many 
theorists
consider to be the best motivated possible extension of the Standard Model
at accessible energies. Supersymmetry predicts the existence of several 
different Higgs bosons, as also discussed in section 2, as well as many 
new supersymmetric particles, the lightest of which may well be produced at 
\LEPTWO. Section 3 discusses the radiative corrections to the 
production of charginos at \LEPTWO, and section 6 discusses the
prospects for slepton detection. 

A novelty at this workshop was the increased interest attracted by
supersymmetric models in which $R$ parity is violated, motivated in
large part by the apparent excess of events at large $Q^2$
observed by the \HONE and \ZEUS collaborations at \HERA~\cite{HERA}, reviewed 
here in section 8, which could
be interpreted as the single production of some squark flavour by an 
$R$-violating Yukawa coupling. This opens up the possibilities of
single leptoquark or squark production at \LEPTWO and of interference effects 
due to the 
exchanges of virtual heavier leptoquarks or squarks, as also mentioned in 
section 8. There is also the possibility of single sneutrino
production, as discussed in section 7. Theoretical upper limits on the 
possible magnitudes of such $R$-violating couplings are discussed in 
section 5 of this report, and other theoretical constraints on 
supersymmetric model building are discussed in section 4.

It is not the purpose of this summary to repeat all the interesting
analyses presented in these previous sections. Rather, I select a few
specific important points that appear to merit further emphasis,
adding further remarks in a few cases. Most of these are related to the
present drive to run \LEP at 200 GeV (`\LEPTWOHUN'), 
if possible during both the 
years 1999 and 2000, or to alternative paradigms for 
supersymmetric phenomenology, including those suggested by 
the \CDF $e^+ e^- \gamma \gamma$ event~\cite{CDF}, as well as by 
interpretations of the \HERA large-$Q^2$ events~\cite{HERA}.

\subsection{Searches for Higgs Bosons}

\begin{figure}
\centerline{\epsfig{figure=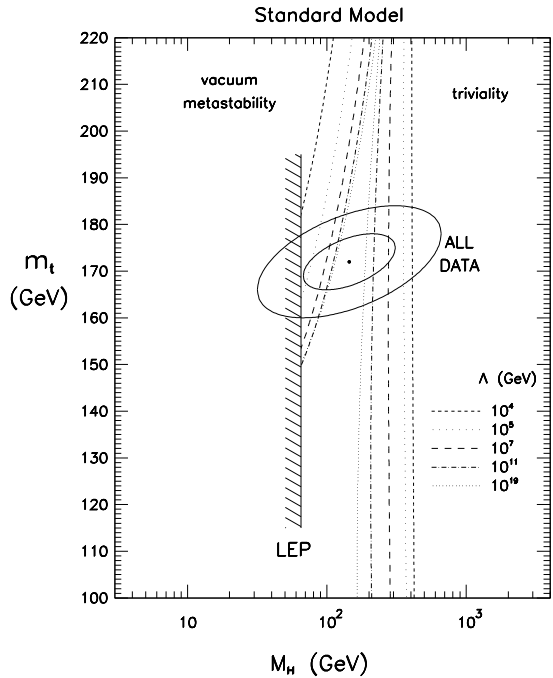,width=7.5cm}}
\caption{
A global fit of $m_H$ to precision electroweak data and Fermilab measurements
indicates a preference for a mass around 140 GeV, with an
uncertainty of a factor of 2~\cite{EFL}, consistent with the validity of the
Standard Model up to a very high scale $\Lambda$, or with
supersymmetry.}
\label{fig:jefig1}
\end{figure}

The search for the Standard Model Higgs boson at \LEPTWO has been
reviewed in section 2. It is worth emphasizing that the precision
electroweak data indicate a preference for a Higgs boson weighing
within a factor of 2 or so of 140 GeV~\cite{EFL}, as seen in 
Fig.~\ref{fig:jefig1}. It is 
also well known 
that masses below about 130 GeV are among the most delicate for the \LHC, 
making the most severe demands on the electromagnetic
calorimetry and/or on $b$ tagging. For these two reasons, 
maximizing the physics reach of \LEPTWO is of capital importance.
It is also important to recall that 
substantial luminosity will be required at $E_{CM} = 200$ GeV if
the discovery potential of \LEPTWOHUN is to be realized fully, because of 
the need to overcome the $Z$ decay background.

As is well known, the minimal supersymmetric extension of the Standard Model
(MSSM) predicts the appearance of a Higgs boson $h$ with mass below about 150
GeV~\cite{MSSMmh}, consistent with the indications from the precision 
electroweak 
data. This upper limit may be reduced to about 100 GeV in models with a
small ratio of Higgs vacuum expectation values $\tan\beta \lappeq$ 2 as
favoured in many theoretical speculations refining the MSSM.
By extending the search region for the Higgs boson beyond 100 GeV,
\LEPTWOHUN would expand significantly the domain of MSSM parameter space accessible
to \LEP.

\begin{figure}
\centerline{\epsfig{figure=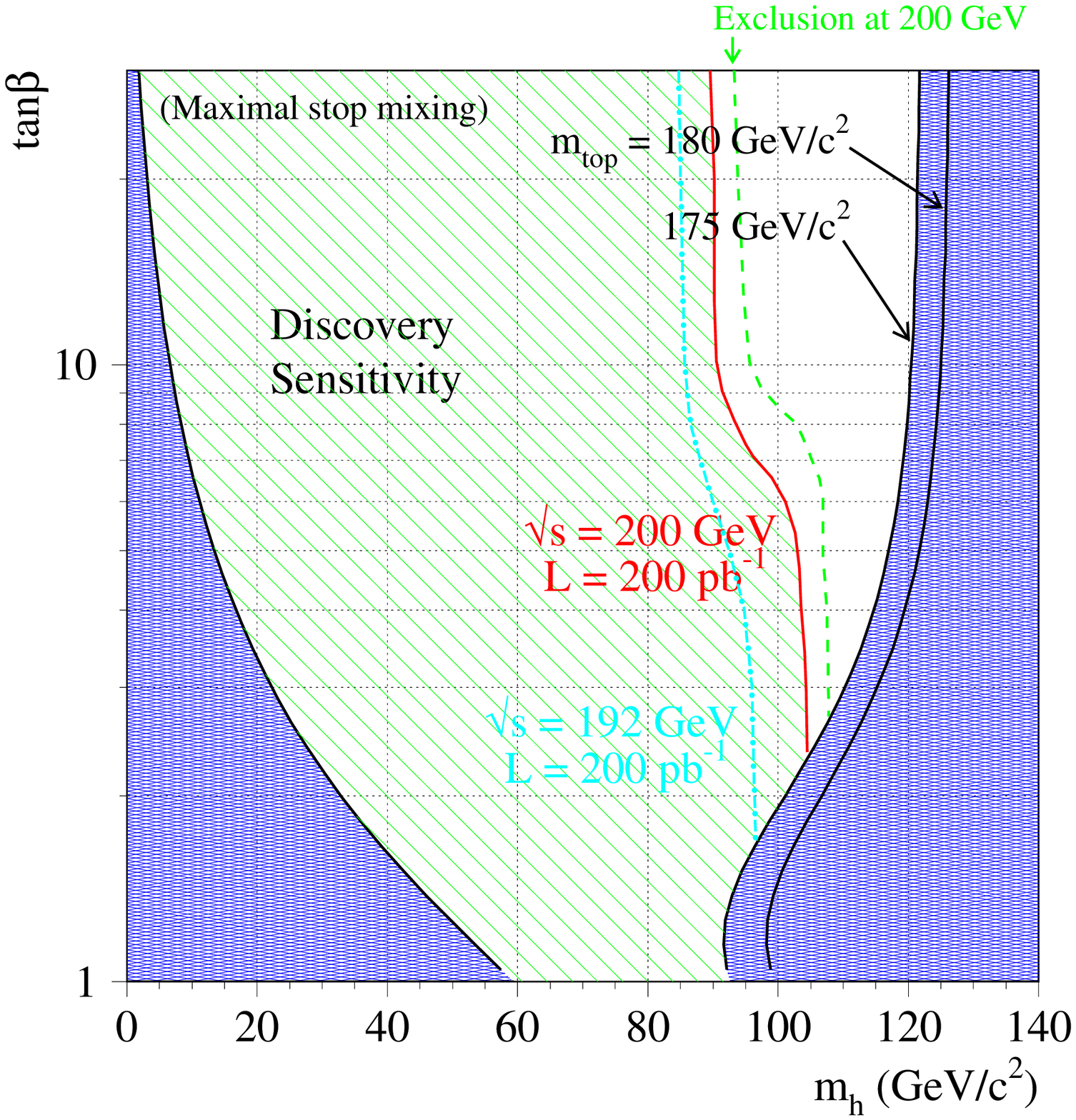,width=7.5cm}}
\caption{
Regions of the $(m_h, {\rm tan}\beta)$ plane characterizing
supersymmetric Higgs bosons, indicating (dark hatching) regions that
are excluded theoretically for different assumed values of the top quark
mass, and regions where supersymmetric Higgs bosons could
be discovered by \LEP running at 192 or 200 GeV, or excluded by running
at 200 GeV~\cite{Janot}.}
\label{fig:jefig2}
\end{figure}

Comparing with Fig.~\ref{fig:MSSM} of section 2 of this report,
Fig.~\ref{fig:jefig2} displays the domains of the MSSM parameters $(m_h, \tan\beta )$ that
are excluded theoretically (dark hatching) and those that could be explored by
 \LEPTWOHUN searches for supersymmetric Higgs bosons (light 
hatching)~\cite{Janot}. 
We see that \LEPTWOHUN would provide significant gains. In particular, 
the lightest Higgs boson could be found with certainty for
$\tan\beta \lappeq 2$, even for unfavourable choices of other MSSM
parameters.

\begin{figure}[htb]
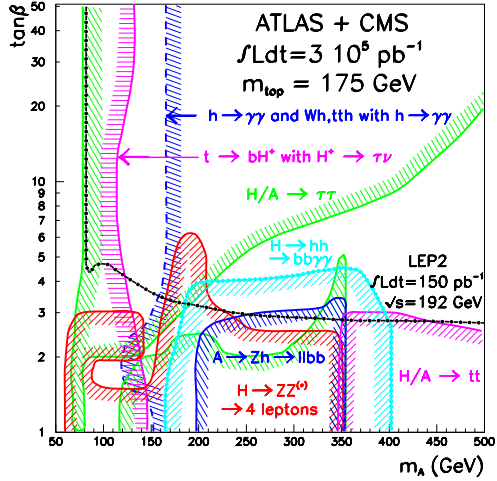

\centerline{ \DESepsf(elzbieta.epsf width 6 cm) }
\caption{
Regions of the $(m_A, \hbox{tan} \beta )$ plane that are accessible
to various searches for MSSM Higgs bosons at the \LHC. Also shown is
the region that could be covered by \LEPTWO at 192 GeV. Increasing
the maximum \LEP energy to 200 GeV would improve significantly the
coverage for tan$\beta \sim 3$ to $4$, where many different \LHC
search strategies come into play~\cite{Was}.}
\label{fig:jefig3}
\end{figure}

The additional coverage available to \LEPTWOHUN
would also complement the \LHC in regions of MSSM parameter space where the
\LHC Higgs searches are particularly delicate and might require combining data,
from both \ATLAS and \CMS, taken over several years of running at the highest
luminosity~\cite{Was}, as seen in Fig.~\ref{fig:jefig3}.

\subsection{Searches for Supersymmetric Particles}

As is well known,
supersymmetry~\cite{susy} is one of the most motivated possible
extensions of the Standard Model - since it may help understand the hierarchy
of mass scales in physics without excessive fine tuning~\cite{hierarchy} - 
and inspires 
many of the current new physics searches at \LEPTWO. In addition, \LEPONE has
provided
two suggestions that supersymmetry may be on the right track. One is the
consistency of the global fit to precise electroweak measurements 
shown in Fig.~\ref{fig:jefig1}
with supersymmetric predictions~\cite{EFL}, and the other
is the consistency of measurements of the gauge coupling strengths with
supersymmetric grand unified theories~\cite{GUT}.

\begin{figure}
\centerline{\epsfig{figure=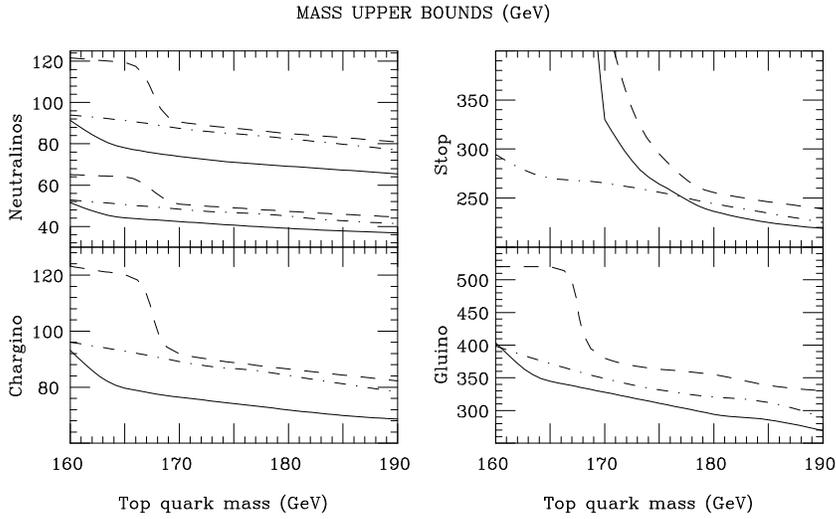,width=11cm}}
\caption{
Upper limits on the masses of various supersymmetric particles
obtained by requiring that $m_Z$ not vary more than 10 times more
rapidly than the input model parameters, under different assumptions on
relations between the latter~\cite{tuning}.}
\label{fig:jefig4}
\end{figure}

Although the full range of possible supersymmetric particle masses can only
be explored using the \LHC, particular interest attaches to masses
around 100 GeV, in the transition
region between \LEP and the \LHC,
where additional \LEPTWO running at the highest possible energy could have
significant impact. This is because the fine tuning required to
stabilize the gauge hierarchy becomes worse for heavier sparticles.
For this reason, there is a preference
for many of  them to be lighter than the maximum of about 2 TeV that can
be explored at the \LHC.
In any given model, the amount of fine tuning can be quantified as the
proportional sensitivity of the $Z$ mass to variations in the input
parameters~\cite{tuning}. As seen in Fig.~\ref{fig:jefig4}, if $m_Z$ is not to vary more 
than 10 times 
more rapidly than the input parameters, the lightest supersymmetric partners of
the electroweak gauge bosons and Higgs bosons - charginos and neutralinos -
are expected to weigh less than about 100 and 60 GeV respectively. \LEPTWOHUN
will enable this range of chargino masses to be explored thoroughly.

\begin{figure}
\centerline{\mbox{\epsfig{figure=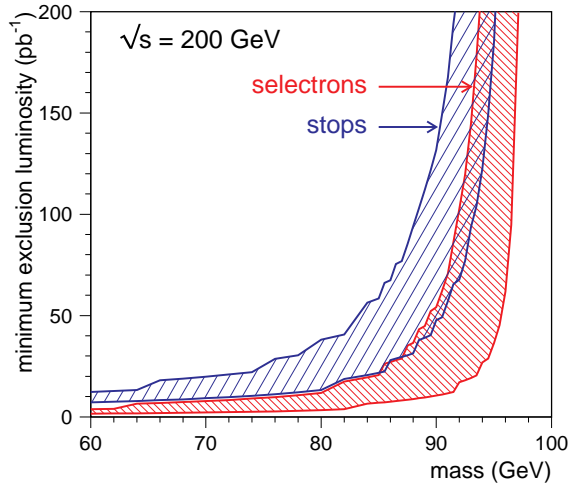,width=7.5cm}}}
\caption{
Domains of the masses of the selectron and lighter stop squark that
could be explored using \LEPTWOHUN, as a function of the total luminosity
obtained~\cite{Schmitt}. The spreads in the curves correspond to differing 
amounts of stop mixing and possible selectron masses.}
\label{fig:jefig5}
\end{figure}

The calculations reported in section 3 will enable the future \LEPTWO
searches for charginos to be interpreted more reliably in terms of
lower limits on the chargino mass: running \LEP at 200 GeV in the
centre of mass would increase the chargino mass reach by about 4 GeV. The 
analysis of section 6
illustrates the gain in smuon mass range that could be attained by
running \LEP at an energy of 200 GeV in the centre of mass. For
example, assuming that the smuon decays into the lightest neutralino, 
and that this is stable, an additional 100 pb$^{-1}$ of integrated 
luminosity at 200 GeV
would enable the physics reach for the smuon mass to be extended by about 
2 GeV.
As a further example in the stable-neutralino context, 
Fig.~\ref{fig:jefig5} shows
that \LEPTWOHUN would  extend the searches for 
the selectron and the lighter stop squark up above 90 GeV~\cite{Schmitt}.
The latter is particularly interesting in the context of 
supersymmetric models that attempt to make a significant
contribution to the $Z$ decay rate into ${\bar b} b$ final states~\cite{Rb}.
Finally, we recall that the lightest neutralino is a prime candidate for the 
cold
dark matter favoured by cosmology and models of structure formation in the
Universe. \LEPTWOHUN would be able to explore completely masses of the lightest
neutralino below about 50 GeV~\cite{EFOS2}, where experiments searching for
astrophysical dark matter have their greatest sensitivity.

In addition to the more complete searches for supersymmetric Higgs bosons 
and other supersymmetric particles in models where the lightest
neutralino is stable, as
reviewed above, \LEPTWOHUN could also settle the
status of models proposed to
explain the \CDF $e^+e^- \gamma \gamma$ event~\cite{CDF}, and cast light on 
possible non-standard interpretations of \HERA large-$Q^2$ data~\cite{HERA},
as discussed in the next two subsections.

\subsection{Search for Supersymmetric  Models with a Light Gravitino}

\begin{figure}
\centerline{\epsfig{figure=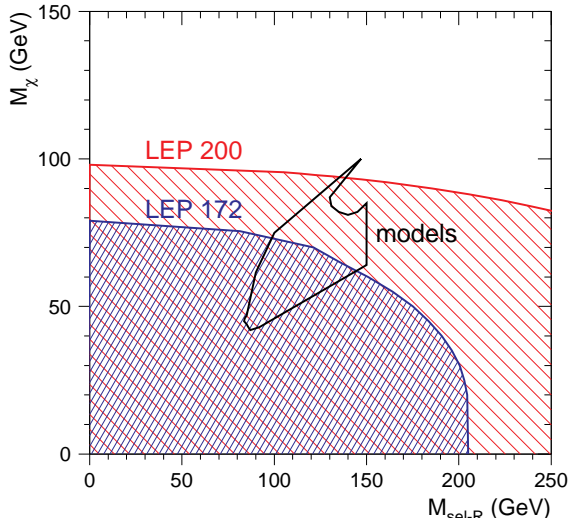,width=7.5cm}}
\caption{
Regions of the slectron - lightest neutralino mass plane that have
already been explored by \LEP and could be explored by \LEPTWOHUN~\cite{ELN}, 
compared with the domain of these parameters postulated 
by explanations of the \CDF $e^+ e^- \gamma \gamma$ event~\cite{CDF} in 
terms of selectron-pair production in models with a light
gravitino.}
\label{fig:jefig6}
\end{figure}

Supersymmetric models with a light gravitino have been the object of 
considerable 
theoretical interest over the last couple of years, motivated in part
by models in which supersymmetry breaking is
communicated to the observable sector by gauge interactions~\cite{messenger},
rather than by gravity. Efforts on these models have
been encouraged~\cite{CDFmodels}  by the \CDF report of an event
containing $e^+ e^- \gamma \gamma$ and missing energy~\cite{CDF}, that 
could be due to radiative decays
of neutralinos into gravitinos. 
Searches at \LEPTWOHUN for events with photons and missing energy have already
excluded a significant fraction of the preferred range of parameters in a
favoured model of this type~\cite{ELN}, as seen in Fig.~\ref{fig:jefig6}.
Furthermore, as 
also seen in
Fig.~\ref{fig:jefig6}, searches at \LEPTWOHUN should be able to explore essentially the 
whole of the
parameter space of this model, and the same is true of alternative
light-gravitino models. Therefore \LEPTWOHUN may be able to deliver a definitive
verdict on this generic scenario for supersymmetry breaking.

\subsection{Search for $R$-Violating Supersymmetry}

In the scenarios discussed in the two previous subsections, it was 
assumed that there is a
multiplicatively conserved quantum number, called $R$ parity, which is
related to baryon and lepton numbers. Another generic possibility is that
although the lightest neutralino is the lightest supersymmetric particle, it
is unstable and decays via new interactions that violate baryon and/or lepton
number and hence $R$ parity. Models of this type have also attracted renewed
attention recently, particularly in connection with possible 
interpretations
of the large-$Q^2$ events seen at \HERA~\cite{HERA} which invoke
the production 
of a squark weighing about 200 GeV by a Yukawa interaction that violates
$R$ parity.

Three specific scenarios of this type have been proposed~\cite{AEGLM}: (a) 
production
of a scharm squark $\tilde c$ off a valence $d$ quark in the proton,
(b) production of the lighter stop squark $\tilde t_1$ off a valence $d$
quark, and (c) production of the $\tilde t_1$ off an $s$ quark in the sea.
One of the important constraints on these scenarios is imposed by the
absence of $e^+ e^- + 2$-jet events at the Tevatron collider~\cite{absence}, 
which requires the $e^+ q$ branching ratio of any such squark to be
significantly less than unity. This occurs in generic domains of the
parameter space for scenario (a), but in only a limited domain of
parameters in scenario (b), with scenario (c) being an
intermediate case~\cite{ELS}.

As was discussed in section 8, the exchange of an $R$-violating
squark could
have measurable effects on $e^+e^-\rightarrow \bar qq$ cross sections at the
highest \LEP energies, even if it cannot be produced directly at \LEP.
However, at the present time this does not appear to rule out even 
scenario (c), which requires the largest Yukawa 
coupling~\footnote{Interesting new bounds on $R$-violating couplings are 
derived in section 5, on the basis of a renormalization-group
analysis of the GUT mass relation $m_b = m_{\tau}$~\cite{CEG}.}, 
and is unlikely ever to challenge scenarios (a) and (b), which require 
much smaller Yukawa couplings. 

As was also mentioned in section 8, there could be an observable 
cross section for
single production of such a quark, via its $R$-violating interactions, if its
mass is below the maximum centre-of-mass energy of \LEP. This possibility puts
a premium on achieving the highest possible energy at \LEP. However, since
all of the \HERA scenarios require $200 \lappeq m_{\tilde q} \lappeq 220$ GeV,
this mechanism may lie for ever beyond the reach of \LEP.

\begin{figure}
\centerline{\epsfig{figure=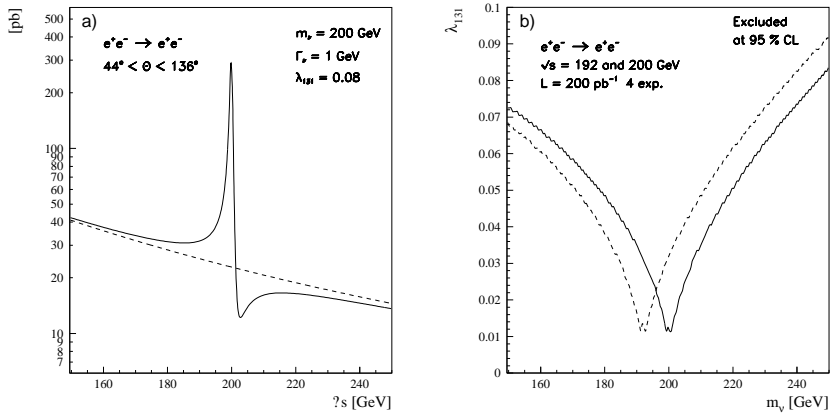,width=11cm}}
\caption{
(a) The possible effect on the $e^+e^-$ annihilation cross
section of a direct-channel sneutrino resonance~\cite{sneutrino}
with an $R$-violating coupling, and (b)
potential limits on such a coupling, as a
function of its mass~\cite{Bourilkov}. It can be seen that \LEPTWOHUN would 
extend significantly the
sensitivity to any direct-channel resonance weighing about 200 GeV.}
\label{fig:jefig7}
\end{figure}

As was discussed in section 7,
there could also be $R$-violating interactions involving just leptons and
their supersymmetric partners. In this scenario, sneutrino exchange
could have measurable effects on
$e^+e^-\rightarrow \ell^+\ell^-$ cross-sections at the highest \LEP 
energies,
and single sneutrino production becomes possible. It
is even possible that \LEPTWOHUN might produce a sneutrino as a direct-channel
resonance. Fig.~\ref{fig:jefig7} shows the possible shape of such a direct-channel
sneutrino resonance~\cite{sneutrino}, as well as the limits 
that \LEP could establish on the coupling
of any such sneutrino, as a function of its mass~\cite{Bourilkov}. We 
see, in particular, the
potential impact of additional running of \LEP at a centre-of-mass energy of
200 GeV.

\subsection{Conclusions}

These examples illustrate the rich area of searches to be
undertaken at \LEPTWO,
in particular in the Higgs sector and in various supersymmetric 
scenarios. Other examples could be given in the
context of composite models, contact interactions, etc.. However, the
capabilities reviewed above 
suffice to indicate that \LEPTWO has every chance to add
discovery of new physics beyond the Standard Model to the
precision $W^+W^-$ and other measurements discussed by other Working 
Groups at this meeting.

In several cases, such as searches for Higgs bosons weighing
around 100 GeV and supersymmetric particles, there is a premium on
running \LEPTWO at $E_{CM} = 200$ GeV for a substantial period,
preferably both the years 1999 and 2000. We therefore urge in the
strongest possible terms that the means to achieve this be made available.



\begin{thebibliography}{99}
%
\bibitem{SM172} 
Barate R \etal, The \ALEPH\ Collaboration, ``Search for the Standard 
  Model Higgs boson in $\ee$ collisions at $\ss =$ 161, 170 and 
  172\,$\G$'', CERN-PPE/97-070, submitted to \PL {\bf B} \\
\noindent
Abreu P \etal, The \DELPHI\ Collaboration, ``Search for neutral 
  and charged Higgs bosons in $\ee$ collisions at $\ss = $ 161\,$\G$ and
  172\,$\G$'', to be submitted to \ZP {\bf C} \\
Acciarri M \etal, The L3 Collaboration, ``Search for the Standard Model 
  Higgs boson in $\ee$ interactions at $161\,\G < \ss < 172\,\G$'', L3 
  preprint 127, to be submitted to \PL {\bf B} \\
Ackerstaff K \etal, The \OPAL\ Collaboration 1997 \PL {\bf B393} 231
%
\bibitem{privcomm} The Joint \LEP\ Higgs Working Group 1997 Private
  communication
%
\bibitem{YellowBook} Accomando E \etal 1996 {\it Physics at \LEPTWO } vol~1 
  ed G Altarelli \etal (Geneva) report CERN 96-01 p~351
%
\bibitem{Chamonix97} Wells P S 1997 {\it Proc. \LEP Performance Workshop
  (Chamonix)} ed J Poole (Geneva) report CERN-SL/97-006 p~137
%
\bibitem{WPHACT} Accomando E and Ballestrero A 1997 {\it Comp. Phys. 
  Comm.} {\bf 99} 230
%
\bibitem{b-tau}
Carena M, Pokorski S and Wagner C E M 1993 \NP {\bf B406} 59
%
\bibitem{mtop}
Barnett R M \etal 1996 \PR {\bf D54} 1 \\
Abe F \etal 1995 \PRL {\bf 74} 2626 \\
Abachi S \etal 1995 \PRL {\bf 74} 2632
%
\end{thebibliography}

\vspace{0.25in}


\subsection{References}

\begin{thebibliography}{99}

\bibitem{MSSMrep}
   H.P. Nilles, \PRTS {\bf 110}, 1 (1984);
   H.E. Haber and G.L. Kane, \PRTS {\bf 117}, 75 (1985);
   R. Barbieri, \RNC {\bf 11}, 1 (1988).

\bibitem{prodLEP}
   H. Komatsu and J. Kubo, \PL {\bf 162B}, 379 (1985);
   R. Barbieri, G. Gamberini, G.F. Giudice, and G. Ridolfi,
   \PL {\bf 195B}, 500 (1987);
   M.S. Carena and C.E.M. Wagner, \PL {\bf 195B},
   599 (1987);
   H. Konig, U. Ellwanger, and M.G. Schmidt, \ZP
   {\bf C36}, 715 (1987);
   A. Bartl, W. Majerotto, and N. Oshimo, \PL
   {\bf B216}, 233 (1989);
   A. Bartl, S. Stippel, W. Majerotto, and N. Oshimo,
   \PL {\bf B233}, 241 (1989);
   A. Bartl, W. Majerotto, N. Oshimo, and S. Stippel,
   \ZP {\bf C47}, 235 (1990);
   K. Hidaka and P. Ratcliffe, \PL {\bf B252},
   476 (1990).

\bibitem{prodLEPII}
   A. Bartl, H. Fraas, and W. Majerotto, \ZP {\bf C30},
   441 (1986);
   A. Bartl, H. Fraas, and W. Majerotto, \NP
   {\bf B278}, 1 (1986);
   A. Bartl, H. Fraas, W. Majerotto, and B. Mosslacher,
   \ZP {\bf C55}, 257 (1992);
   M.H. Nous, M. El-Kishen, and T.A. El-Azem, \MPL
   {\bf A7}, 1535 (1992);
   J. Feng and M. Strassler, \PR {\bf D51}, 4661 (1995); 
   A.S. Belyaev and A.V. Gladyshev, Report No. JINR-E2-97-76, Mar. 1997.

\bibitem{others}
   H. Baer and X. Tata, Report No. FSU-HEP-921222, Dec. 1992, published 
   in Erice 1992, 23rd Eloisatron workshop, Properties of SUSY particles, 
   p. 244-271;\\
   J.L. Feng and M.J. Strassler, \PR {\bf D55}, 1326 
   (1997);\\
   M.A. Diaz, \MPL {\bf A12}, 307 (1997).

\bibitem{us}
   M.A. Diaz and S.F.King, \PL {\bf B349}, 105 (1995); \\
   M.A. Diaz and S.F.King, \PL {\bf B373}, 100 (1996).

\bibitem{expcha}
   D. Buskulic \etal, \ALEPH Collaboration, \PL {\bf B373},
   246 (1996);\\ D. Buskulic \etal, ALEPH Collaboration, 
   \PL {\bf B384}, 461 (1996);\\
   P. Abreu \etal, DELPHI Collaboration, \PL {\bf B382},
   323 (1996);\\
   M. Acciarri \etal, \L3 Collaboration, \PL {\bf B377},
   289 (1996);\\
   G. Alexander \etal, \OPAL Collaboration, \PL {\bf B377},
   181 (1996);\\ K. Ackerstaff \etal, OPAL Collaboration, 
   \PL {\bf B389}, 616 (1996);\\ G. Alexander \etal, OPAL Collaboration, 
   \ZP {\bf C73}, 201 (1997).

\bibitem{pierce}
   A.B. Lahanas, K. Tamvakis, and N.D. Tracas,  
   \PL {\bf B324}, 387 (1994);\\
   D. Pierce, A. Papadopoulos,\PR {\bf D50}, 565 (1994),
   and \NP {\bf B430}, 278 (1994).

\bibitem{dkr}
M.A. Diaz, S.F. King, and D.A. Ross, in preparation.

\end{thebibliography}

\subsection{References}

\begin{thebibliography}{99}
\bibitem{CCB}
J.~M.~Frere, D.~R.~T.~Jones and S.~Raby, 
\NP {\bf B222}, 11 (1983);\\ 
L.~Alvarez-Gaume, J.~Polchinski and M.~Wise,
\NP {\bf B221}, 495 (1983);\\ 
M.~Drees, M.~Gluck and K.~Grassie, 
\PL {\bf B157}, 164 (1985);\\ J.~F.~Gunion, H.~E.~Haber and M.~Sher,
\NP  {\bf B306}, 1 (1988);\\ P.~Langacker and N.~Polonsky,
\PR {\bf D 50}, 2199 (1994);\\   
A.~J.~Bordner, KUNS-1351 (hep-ph/9506409);\\
A. Strumia, \NP {\bf B482} (1996) 24.
\bibitem{sash1}
A. Kusenko, P. Langacker and G. Segre, \PR {\bf D54} (1996) 5824;\\
A. Kusenko, \NPPS {\bf 52A} (1997) 67.
\bibitem{callmun}
J.A. Casas, A. Lleyda and C. Munoz, \NP {\bf B471} (1996) 3.
\bibitem{UFB}
H. Komatsu, \PL {\bf B215} (1988) 323.
\end{thebibliography}

\subsection{References}

\begin{thebibliography}{99}
\bibitem{bhatt}
G. Bhattacharyya, \NPPS {\bf 52A} (1997) 83.
\bibitem{guts}
P. Langacker and N. Polonsky, \PR {\bf D50} (1994) 2199.
\bibitem{lang}
P. Langacker and N. Polonsky, \PR {\bf D47} (1993) 4028.
\bibitem{thresh}
D.M. Pierce, J.A. Bagger, K.T. Matchev and R.J. Zhang,
\NP {\bf B491} (1997) 3.
\bibitem{us2}
B.C. Allanach, H. Dreiner and H. Pois, work in progress.
\end{thebibliography}

\subsection{References}
\begin{thebibliography}{1}
\bibitem{direct}
J. Erler, J.L. Feng, N. Polonsky, \PRL {\bf 78} (1997) 3063;
J. Kalinowski et. al., hep-ph/9703436
\bibitem{WW}
C.F. Weisz\"{a}cker, \ZP {\bf 88} (1934) 612; E.J. Williams, \PR {\bf
45} (1934) 729
\bibitem{Frixione}
S. Frixione et. al. \PL {\bf B319} (1993) 339
\bibitem{bhatt2}
V. Barger, G. F. Giudice and T. Han, \PR {\bf D40} (1989) 2987;
G. Bhattacharyya, \NPPS {\bf 52A} (1997) 83;
H. Dreiner, hep-ph/9707435
\end{thebibliography}

\subsection{References}

\begin{thebibliography}{99}
\bibitem{H1}
 \HONE Collab., C. Adloff et al., Z. Phys. C74 (1997) 191.
\bibitem{ZEUS}
 \ZEUS Collab., J. Breitweg et al., Z. Phys. C74 (1997) 207.
\bibitem{Hewett}
J.L. Hewett and T.G. Rizzo, hep-ph/970337.
\bibitem{preprints}
P.H. Frampton, hep-ph/9706220 from June 1997
catalogues 30 related preprints from early 1997. 
\bibitem{Ellis}
G. Altarelli et al., hep-ph/9703276.
\bibitem{relevant}
H. Dreiner and P. Morawitz, hep-ph/9703279.
J. Kalinowski et al., hep-ph/9703288.
C.G. Papadopoulos, hep-ph/9703372.
S. Jadach, B.F.L. Ward and Z. Was, hep-ph/9704241.
\bibitem{Drees}
M.Drees, hep-ph/9703332.
\bibitem{Doncheski}
M.A. Doncheski and S. Godfrey, Phys. Lett. B393(1997)355.
M.A. Doncheski and S. Godfrey, hep-ph/9703285.
\bibitem{YR}
S. Ambrosanio et al. in CERN 96-01, Vol 1, Ed. G. Altarelli et al, pp506-510.
\bibitem{D0}
D0 Collab., B. Abbott et al, hep-ex/9707033.
\bibitem{CDF1}
CDF. COllab., F. Abe et al, Phys. Rev. Lett 78 (1997) 2906.
\bibitem{SSR}
S. Soeldner-Rembold, hep-ex/9706003.
\bibitem{OPAL_161}
\OPAL\ Collab., K. Ackerstaff et al., Phys. Lett. B391 (1997) 221.
\bibitem{HERE}
These proceedings.
\end{thebibliography}

\subsection{References}


\begin{thebibliography}{99}


\bibitem{LEP2YB} {\it Proceedings of the Workshop on Physics at \LEPTWO}, eds.
G. Altarelli, T.~Sj\"ostrand and F. Zwirner, CERN Report 96-01.

\bibitem{ZH} J. Ellis, M.K. Gaillard and D.V. Nanopoulos, \NP {\bf B106} (1976)
292;\\ B.L. Ioffe and V.A. Khoze, {\it Sov.J.Part.Nucl.} {\bf 9} (1978) 50;\\
B.W. Lee, C. Quigg and H. Thacker, \PRL {\bf 38} (1977) 883 and \PR {\bf D16} (1977)
1519.

\bibitem{susy} Y.A. Gol'fand and E.P. Likhtman, {\it Pis'ma Zh.E.T.F.} {\bf 13} (1971)
323;\\
D. Volkov and V.P. Akulov, \PL {\bf 46B} (1973) 109;\\
J. Wess and B. Zumino, \NP {\bf B70} (1974) 39;\\
for a review, see P. Fayet and S. Ferrara, \PRTS {\bf
32C} (1977) 249.

\bibitem{HERA} \HONE Collaboration, \ZP {\bf C74} (1997) 191;\\
\ZEUS Collaboration, \ZP {\bf C74} (1997) 207.

\bibitem{CDF} S. Park, in {\it Proceedings of the 10th Topical Workshop on
Proton-Antiproton Collider Physics}, Fermilab, 1995, eds. R. Raja and J. Yoh
(AIP, New York, 1995), p. 62.

\bibitem{EFL} \LEP Electroweak Working Group, CERN preprint PPE/97-183;\\
J. Ellis, G.L. Fogli and E. Lisi, \PL {\bf B389} (1996) 321.

\bibitem{MSSMmh} Y. Okada, M. Yamaguchi and T. Yanagida, \PTP {\bf 85} 
(1991) 1;\\
J. Ellis, G. Ridolfi and F. Zwirner, \PL {\bf B257} (1991) 83 and \PL {\bf 
B262} (1991) 477;\\
H.E. Haber and R. Hempfling, \PRL {\bf 66} (1991) 1815;\\
R. Barbieri, M. Frigeni and F. Caravaglios, \PL {\bf B258} (1991) 167;\\
Y. Okada, M. Yamaguchi and T. Yanagida, \PL {\bf B262} (1991) 54.

\bibitem{Janot} P. Janot, private communication.

\bibitem{Was} E. Richter-Was, D. Froidevaux, F. Gianotti, L. Poggioli, 
D. Cavalli  and S. Resconi, CERN Preprint TH/96-111 (1996).

\bibitem{hierarchy} L. Maiani, {\it Proceedings Summer 
School on Particle Physics},
Gif-sur-Yvette, 1979 (IN2P3, Paris, 1980), p. 3;\\
G. 't Hooft, in {\it Recent Developments in Field Theories}, eds. G. 't
Hooft et al. (Plenum Press, New York, 1980);\\
E. Witten, \NP {\bf B188} (1981) 513;\\
R.K. Kaul, \PL {\bf 109B} (1982) 19.

\bibitem{GUT} S. Dimopoulos, S. Raby and F. Wilczek, \PR {\bf D24}
(1981) 1681;\\
W.J. Marciano and G. Senjanovic, \PR {\bf D25} (1982) 3092;\\
L.E. Ib\`a\~nez and G.G. Ross, \PL {\bf 105B} (1982) 439;\\
M.B. Einhorn and D.R.T. Jones, \NP {\bf B196} (1982) 475;\\
J. Ellis, S. Kelley and D.V. Nanopoulos, \PL {\bf B249} (1990) 441 
and {\bf B260} (1991) 131;\\
P. Langacker and M. Luo, \PR {\bf D44} (1991) 817;\\
U. Amaldi, W. de Boer and H. Furstenau, \PL {\bf B260} (1991) 447;\\
F. Anselmo, L. Cifarelli, A. Petermann and A. Zichichi, \NC {\bf  104A} 
(1991) 1817.

\bibitem{tuning}J. Ellis, K. Enqvist, D.V. Nanopoulos and F. Zwirner, 
\MPL {\bf A1} (1986) 57;\\
R. Barbieri and G.F. Giudice, \NP {\bf B306} (1988) 63;\\
S. Dimopoulos and G.F. Giudice, \PL {\bf B357} (1995) 573.

\bibitem{Schmitt} M. Schmitt, private communication.

\bibitem{Rb} A. Djouadi et al., \NP {\bf B349} (1991) 48;\\
M. Boulware and D. Finell, \PR {\bf D44} (1991) 2054;\\
G. Altarelli, R. Barbieri and F. Caravaglios, \PL {\bf B314} (1993) 357;\\
D. Garcia and J. Sola, \PL {\bf B357} (1995) 349;\\
X. Wang, J. Lopez and D.V. Nanopoulos, \PR {\bf D52} (1995) 4116;\\
M. Shifman, \MPL {\bf A10} (1995) 605;\\
G.L. Kane, R.G. Stuart and J.D. Wells, \PL {\bf B354} (1995) 350;\\
J. Erler and P. Langacker, \PR {\bf D52} (1995) 441;\\
P.H. Chankowski and S. Pokorski, \NP {\bf B475} (1996) 3;\\
J. Ellis, J.J. Lopez and D.V. Nanopoulos, \PL {\bf B372} (1996) 95
and \PL {\bf B397} (1997) 88.

\bibitem{EFOS2} J. Ellis, T. Falk, K.A. Olive and M. Schmitt,
CERN preprint TH/97-105, hep-ph/9705444.

\bibitem{messenger} See, e.g., M. Dine and A. Nelson, \PR {\bf D48}
(1993) 1277, {\bf D51} (1995) 1362 and {\bf D53} (1996) 2658.

\bibitem{CDFmodels} D. Stump, M. Wiest and C.-P. Yuan, \PR {\bf D54}
(1996) 1936;\\
S. Dimopoulos, M. Dine, A. Raby and S. Thomas, \PRL {\bf 76} (1996) 3494;\\
S. Ambrosanio, G. Kane, G. Kribs, S. Martin and S. Mrenna, \PRL {\bf 76} (1996)
3498 and \PR {\bf D54} (1996) 5395;\\
S. Dimopoulos, S. Thomas and J. Wells, \PR {\bf D54} (1996) 3283;\\
K. Babu, C. Kolda and F. Wilczek, \PRL {\bf 77} (1996) 3070;\\
J.L. Lopez and D.V. Nanopoulos, {\it Mod. Phys. Lett.} {\bf A10} (1996)
2473 and \PR {\bf D55} (1997) 4450;\\
J.L. Lopez, D.V. Nanopoulos and A. Zichichi, \PRL {\bf 77} (1996) 5168
and \PR {\bf D55} (1997) 5813.

\bibitem{ELN} J. Ellis, J.L. Lopez and D.V. Nanopoulos, \PL {\bf
B394} (1997) 354.

\bibitem{AEGLM}
J. Butterworth and H. Dreiner, \NP {\bf B397} (1993) 3;\\
H. Dreiner and P. Morawitz, \NP {\bf B428} (1994) 31;\\
E. Perez, Y. Sirois and H. Dreiner, Contribution to Beyond the Standard Model
Group, 1995-1996 Workshop on Future Physics at \HERA, see also the Summary by H.
Dreiner, H.U. Martyn, S. Ritz and D. Wyler, hep-ph/9610232;\\
T. Kon and T. Kobayashi, \PL {\bf B270} (1991) 81;\\
T. Kon, T. Kobayashi and S. Kitamura, \PL {\bf B333} (1994) 263;\\
T. Kon, T. Kobayashi, S. Kitamura, K. Nakamura and S. Adachi, {\it Z.Phys.} {\bf
C61} (1994) 239;\\
T.. Kobayashi, S. Kitamura and T. Kon, \IJMP {\bf A11} (1996) 1875;\\
D. Choudhury and S. Raychaudhuri, \PL {\bf B401} (1997) 54;\\
G. Altarelli, J. Ellis, G.F. Giudice, S. 
Lola and M.L. Mangano, CERN preprint TH/97-40, hep-ph/9703276.



\bibitem{absence} H.S. Kambara, for the \CDF Collaboration, hep-ex/9706026;\\
D0 collaboration, B. Abbott et al., hep-ex/9707033.

\bibitem{ELS} J. Ellis, S. Lola and K. Sridhar, CERN preprint TH/97-109,
hep-ph/9705416.

\bibitem{CEG} M.S. Chanowitz, J. Ellis and M.K. Gaillard, \NP {\bf B128}
(1977) 506.

\bibitem{sneutrino} J. Kalinowski, R. R\"uckl, H. Spiesberger and P.
Zerwas, hep-ph/9703436.

\bibitem{Bourilkov} D. Bourilkov, private communication.

\end{thebibliography}
\end{document}